\documentclass[a4paper,preprint,pra]{revtex4}

\usepackage{graphicx}
\usepackage{dcolumn}
\usepackage{bm}
\usepackage{color}

\newcommand{\eqa}{\begin{equation}}
\newcommand{\eqz}{\end{equation}}
\newcommand{\eqma}{\begin{eqnarray}}
\newcommand{\eqmz}{\end{eqnarray}}

\begin{document}
\newcommand{\e}{{\em e}~}
\title{Benchmark thermochemistry of the C$_n$H$_{2n+2}$ alkane isomers (n=2--8) and performance of DFT and composite ab initio methods for dispersion-driven isomeric equilibria\footnote{Dedicated to Prof. Yitz\d{h}ak Apeloig on the occasion of his 65th birthday}}
\author{Amir Karton, David Gruzman, and Jan M. L. Martin*}
\affiliation{Department of Organic Chemistry, 
Weizmann Institute of Science, 
IL-76100 Re\d{h}ovot, Israel}
\email{gershom@weizmann.ac.il}

\date{{\em J. Phys. Chem. A} MS {\bf  jp-2009-04369h}: Received May 11, 2009; Accepted May 15, 2009}

\begin{abstract}
The thermochemistry of linear and branched alkanes with up to eight carbons has been reexamined by means of W4, W3.2lite and W1h theories. `Quasi-W4' atomization energies have been obtained via isodesmic and hypohomodesmotic reactions. Our best atomization energies at 0 K (in kcal/mol) are: 1220.04 $n$-butane, 1497.01 $n$-pentane, 1774.15 $n$-hexane, 2051.17 $n$-heptane, 2328.30 $n$-octane, 1221.73 isobutane, 1498.27 isopentane, 1501.01 neopentane, 1775.22 isohexane, 1774.61 3-methylpentane, 1775.67 diisopropyl, 1777.27 neohexane, 2052.43 isoheptane, 2054.41 neoheptane, 2330.67 isooctane, and 2330.81 hexamethylethane. Our best estimates for $\Delta H^\circ_{f,298K}$ are: -30.00 $n$-butane, -34.84 $n$-pentane, -39.84 $n$-hexane, -44.74 $n$-heptane, -49.71 $n$-octane, -32.01 isobutane, -36.49 isopentane, -39.69 neopentane, -41.42 isohexane, -40.72 3-methylpentane, -42.08 diisopropyl, -43.77 neohexane, -46.43 isoheptane, -48.84 neoheptane, -53.29 isooctane, and -53.68 hexamethylethane. These are in excellent agreement (typically better than 1 kJ/mol) with the experimental heats of formation at 298 K obtained from the CCCBDB and/or NIST Chemistry WebBook databases. However, at 0 K a large discrepancy between theory and experiment (1.1 kcal/mol) is observed for only neopentane. This deviation is mainly due to the erroneous heat content function for neopentane used in calculating the 0 K CCCBDB value. The thermochemistry of these systems, especially of the larger alkanes, is an extremely difficult test for density functional methods. {\em A posteriori} corrections for dispersion are essential. Particularly for the atomization energies, the B2GP-PLYP and B2K-PLYP double-hybrids, and the PW6B95 hybrid-meta GGA clearly outperform other DFT functionals. 
\end{abstract}

\maketitle

\section{Introduction\label{sec:Int}}

The fundamental importance of alkanes as organic chemistry building blocks and in industrial chemistry (particularly petrochemistry) is self-evident to any chemist.

Linear and branched lower alkanes are the principal components of gaseous and liquid fossil fuels. Accurate knowledge of their
thermodynamic properties is essential for reliable computational modeling of combustion processes. (We note that one of us, JMLM, is a member of a IUPAC task group working in this area.\cite{iupac1})

Aside from their practical relevance, alkanes present some intriguing methodological issues. The importance of accurate zero-point vibrational energies and diagonal Born-Oppenheimer corrections has been discussed previously\cite{C2H6}, and this applies to both wavefunction ab initio and density functional methods. While post-CCSD(T) computational thermochemistry methods like W4 theory\cite{w4,w4.4} or HEAT\cite{HEAT1,HEAT2,HEAT3} have no trouble dealing with systems that, from an electronic structure 
point of view, are much more taxing than alkanes, their steep cost scaling makes application to higher alkanes (or higher hydrocarbons in general) impractical at present.

Density functional theory seems to be the obvious alternative. However, in recent years a number of authors\cite{Yao03,Gilbert05,Radom05,Grimme06,Gri07,Schle06,Schle06b,Schre07} have pointed to a disturbing phenomenon\cite{RefRep}: the error in computed atomization energies of $n$-alkanes grows in direct proportion to the chain length. In addition, these same authors found that popular DFT methods have significant problems with hydrocarbon isomerization energies in general, and alkane isomerization energies in particular. This latter problem appears to be related to the poor description of dispersion by most DFT functionals and can be remedied to a large extent by empirical dispersion corrections\cite{Grimme06}.

For molecules as chemically systematic as alkanes, a computationally more cost-effective approach than brute-force atomization energy calculations is the use of bond separation reactions, such as isodesmic\cite{Isodesmic} and homodesmotic\cite{Homodesmotic} reactions. Recently, Schleyer and coworkers\cite{Schleyer08, Schleyer09} discussed the concepts of `protobranching' and of `hypohomodesmotic reactions', i.e., reactions which, in addition to being isodesmic (i.e., conserving numbers of each formal bond type),  conserve the number of C atoms in each hybridization state and hapticity (primary, secondary, tertiary, quaternary). The latter is a refinement of the earlier `homodesmotic reaction' concept\cite{Homodesmotic}.

They established a consistent hierarchy of hydrocarbon reaction types that successively conserve larger molecular fragments: atomization $\supseteq$ isogyric $\supseteq$ isodesmic $\supseteq$ hypohomodesmotic $\supseteq$ homodesmotic $\supseteq$ hyperhomodesmotic which provides a converging sequence in the sense that the energetic components of the reaction cancels to a larger extent between reactants and products as the reaction hierarchy is traversed. 

In the present work we obtain `quasi-W4' atomization energies for C$_4$--C$_8$ alkanes through the use of isodesmic and hypohomodesmotic reaction cycles that involve only methane, ethane, and propane in addition to one larger alkane. The reaction energies are calculated at the W3.2lite or W1h levels, while for methane, ethane, and propane W4 benchmark values are used. We shall show that the reaction energies of hypohomodesmotic reactions and judiciously selected isodesmic reactions are well converged even at the W1h level. The use of hypohomodesmotic reactions leads to near-perfect cancellation of valence correlation effects, and the use of judiciously selected isodesmic reactions leads to near-perfect cancellation of post-CCSD(T) correlation effects. 

We will then proceed to evaluate a number of DFT functionals and composite ab initio thermochemistry methods against the reference values obtained, both for the atomization and for the isomerization energies.

\section{Computational methods\label{sec:Com}}
All calculations were carried out on the Linux cluster of the Martin group at Weizmann. DFT geometry optimizations were carried out using Gaussian 03 Revision E.01\cite{g03}. The B3LYP\cite{LYPc,B3,B3LYP} DFT hybrid exchange-correlation (XC) functional was used in conjunction with the pc-2\cite{pc-2} polarization consistent basis set of Jensen. All large-scale self-consistent field (SCF), CCSD and CCSD(T) calculations\cite{Rag89,Wat93} were carried out with the correlation consistent family of Dunning and coworkers\cite{Dun89,Ken92,Wil01,pwCVnZ,Dyall} using version 2006.1 of the MOLPRO\cite{molpro} program system. All single-point post-CCSD(T) calculations were carried out using an OpenMP-parallel version of M. K\'allay's general coupled cluster code MRCC\cite{mrcc} interfaced to the Austin-Mainz-Budapest version of the ACES II program system\cite{aces2de}. The diagonal Born-Oppenheimer correction (DBOC) calculations were carried out using the CFOUR program system\cite{cfour}.

The computational protocols of W{\it n}h theories W1\cite{w1,w1h}, W3.2lite\cite{w3.2lite}, and W4\cite{w4} used in the present study have been specified and rationalized in great detail elsewhere\cite{w1,w1h,w4,w3.2lite}. (Throughout, W3.2lite refers to variant W3.2lite(c) as described in ref. \cite{w3.2lite}.) The use of the W{\em n}h variants of the W{\em n} methods, in which the diffuse functions are omitted from carbon and less electronegative elements, is of no thermochemical consequence for neutral alkanes\cite{w1h}, but computer resource requirements are substantially reduced.

For the sake of making the paper self-contained, we will briefly outline the various steps in W3.2lite theory and in W4h theory for first-row elements:
\begin{itemize}
\setlength{\itemsep}{0in}
\item reference geometry and ZPVE correction are obtained at the B3LYP/pc-2 level of theory for W3.2lite, and at the 
CCSD(T)/cc-pVQZ level for W4h. 
\item the ROHF-SCF contribution is extrapolated using the Karton-Martin modification\cite{Kar-Mar} of Jensen's extrapolation formula\cite{Jen05}:
\begin{equation}
E_{{\rm HF},L}=E_{{\rm HF},\infty}+A(L+1)\exp(-9\sqrt{L})
\end{equation}
For W3.2lite and W4h, the extrapolations are done from the cc-pV\{Q,5\}Z and cc-pV\{5,6\}Z basis set pairs, respectively.
\item the RCCSD valence correlation energy is extrapolated from these same basis sets.
Following the suggestion of Klopper\cite{Klop01}, $E_{\rm corr,RCCSD}$ is partitioned in singlet-coupled pair energies, triplet-coupled pair energies, and $\hat{T}_1$ terms. The $\hat{T}_1$ term (which exhibits very weak basis set dependence) is simply set equal to that in the largest basis set, while the singlet-coupled and triplet-coupled pair energies are extrapolated using $A+B/L^\alpha$ with $\alpha_S$=3 and $\alpha_T$=5. 
\item the (T) valence correlation energy is extrapolated from the cc-pV\{T,Q\}Z basis set pair for W3.2lite, and from cc-pV\{Q,5\}Z for W4h. For open-shell systems, the Werner-Knowles-Hampel (a.k.a., MOLPRO) definition\cite{Ham2000} of the restricted open-shell CCSD(T) energy is employed throughout, rather than the original Watts-Gauss-Bartlett\cite{Wat93} (a.k.a. ACES II) definition.
\item the CCSDT$-$CCSD(T) difference, $\hat{T_3}-$(T), in W3.2lite is obtained from the empirical expression 2.6$\times$cc-pVTZ(no $f$ 1$d$)(no $p$ on H)--1.6$\times$cc-pVDZ(no $p$ on H), where the CCSDT energy is calculated using ACES II. In W4h it is instead extrapolated using $A+B/L^3$ from cc-pV\{D,T\}Z basis sets.
\item the difference between ACES II and MOLPRO definitions of the valence RCCSD(T) definition is extrapolated from cc-pVDZ and cc-pVTZ basis sets. One-half of this contribution is added to the final result, as discussed in the appendix of ref. \cite{w4}.
\item post-CCSDT contributions in W3.2lite are estimated from UCCSDT(Q)/cc-pVDZ(no $p$ on H)--UCCSDT/cc-pVDZ(no $p$ on H) scaled by 1.1. In W4h, the connected quadruples are obtained as 1.1$\times$[UCCSDT(Q)/cc-pVTZ $-$ UCCSDT/cc-pVTZ + UCCSDTQ/cc-pVDZ $-$ UCCSDT(Q)/cc-pVDZ], while the contribution of connected quintuple excitations is evaluated at the CCSDTQ5/cc-pVDZ(no $d$) level.
\item the inner-shell correlation contribution, in both cases, is extrapolated from RCCSD(T)/cc-pwCVTZ and RCCSD(T)/cc-pwCVQZ calculations.
\item the scalar relativistic contribution, again in both cases, is obtained from the difference between nonrelativistic RCCSD(T)/cc-pVQZ and second-order Douglas-Kroll RCCSD(T)/DK-cc-pVQZ calculations. 
\item atomic spin-orbit coupling terms are taken from the experimental fine structure.
\item finally, a diagonal Born-Oppenheimer correction (DBOC) is obtained at the ROHF/cc-pVTZ level.
\end{itemize}
The main changes in W1h relative to W3.2lite are: (a) the SCF component is extrapolated from the cc-pV\{T,Q\}Z basis sets, using the formula $A+B/L^5$; (b) The valence RCCSD component is extrapolated from the same basis sets, using $A+B/L^{3.22}$; (c) the valence parenthetical triples, (T), component is extrapolated from cc-pV\{D,T\}Z basis sets, using $A+B/L^{3.22}$; (d) inner-shell correlation contributions are evaluated at the CCSD(T)/MTsmall level; and (e) post-CCSD(T) correlation effects as well as the DBOC are completely neglected. 

The CCSDTQ5/cc-pVDZ(no $d$) calculation for propane proved to be too taxing even for our strongest machine (8-core, Intel Cloverton 2.66 GHz, with 32 GB of RAM). For the alkanes for which we do have this term, CH$_4$ and C$_2$H$_6$, it is practically zero (0.00 and 0.01 kcal/mol, respectively), therefore for propane it was safely neglected. 

The anharmonic zero-point vibrational energy (ZPVE) of propane, propene, propyne, and allene was calculated using the following equation\cite{AllenZPVE}, 
\begin{equation}
{\rm ZPVE}= \frac{1}{2}\sum_i{\omega_i}-\frac{1}{32}\sum_{ijk}{\frac{\phi_{iik}\phi_{kjj}}{\omega_k}}-\frac{1}{48}\sum_{ijk}{\frac{\phi_{ijk}^2}{\omega_i+\omega_j+\omega_k}}+\frac{1}{32}\sum_{ij}{\phi_{iijj}}+{\rm Z}_{kinetic}, 
\end{equation}
where the cubic, quartic, and kinetic energy terms were computed at the MP2/cc-pVTZ level of theory, and the harmonic term was partitioned into valence and core-valence contributions which were calculated at the CCSD(T)/cc-pVQZ and CCSD(T)/MTsmall levels of theory, respectively. (For propane we resorted to a CCSD(T)/cc-pVTZ calculation since the CCSD(T)/cc-pVQZ proved too daunting; based on the results for the other systems this is expected to have little effect, e.g., the differences between the harmonic ZPVE calculated with the two basis sets are: 0.02, 0.02, 0.02, and 0.01 kcal/mol for methane, ethane, propene, and allene, respectively.) 

Unless noted otherwise, experimental data for the heats of formation at 0 K were taken from the NIST Computational Chemistry Comparison and Benchmark Database (CCCBDB)\cite{cccbdb}. The atomization energies quoted in CCCBDB assume CODATA\cite{codata} values for the atomic heats of formation at 0 K: however, particularly for carbon atom, the ATcT value\cite{Branko-carbon} (170.055$\pm$0.026 kcal/mol) is significantly higher than the CODATA value (169.98$\pm$0.11 kcal/mol), consequently, using the ATcT value in converting $\Delta H^\circ_{f,0K}$ to atomization energy raises the atomization energy over the CCCBDB value by $m\times 0.075$ kcal/mol for a system with $m$ carbon atoms. Thus, throughout the paper, the experimental TAE$_0$ were obtained from the heats of formation at 0 K using ATcT atomic heats of formation at 0 K (C 170.055$\pm$0.026, and H 51.633$\pm$0.000 kcal/mol)\cite{Branko-carbon}. 
In cases where only experimental heats of formation at 298 K are available from CCCBDB ($n$-heptane, $n$-octane, isoheptane, and isooctane), they were first converted to 0 K using $H_{298}-H_0$ for H$_2$(g) 2.024$\pm$0.000, and C(cr,graphite) 0.251$\pm$0.005 kcal/mol from CODATA\cite{codata}, and the molecular heat content functions from the TRC (Thermodynamic Research Center) tables\cite{TRC} which are the source of virtually all CCCBDB enthalpy functions for the species considered in the present paper.

In order to facilitate direct comparison with experiment, we have also converted our calculated atomization energies to heats of formation at room temperature, $\Delta H^\circ_{f,298K}$. 
Rather than mix our calculated atomization energies with the TRC enthalpy functions, we have calculated our own $H_{298}-H_0$ for the alkanes. 
The translational, rotational, and vibrational contributions were obtained within the RRHO (rigid rotor-harmonic oscillator) approximation from the B3LYP/pc-2 calculated geometry and harmonic frequencies. Internal rotation corrections were obtained using the Ayala-Schlegel method\cite{AyalaSchlegel}, again on the B3LYP/pc-2 potential surface. This leaves us with the issue of correcting for the ensemble of low-lying conformers of the alkanes: by way of illustration, $n$-butane through $n$-octane have 2, 4, 12, 30, and 96 unique conformers, respectively. The relative energies of these conformers (which are surprisingly sensitive to the level of theory as they are strongly driven by dispersion) were the subject of a recent benchmark study by our group\cite{Conformers}. While large basis set CCSD(T) calculations for all conformers of the heptanes and octanes proved too costly (primarily for those without any symmetry), it was found in Ref.\cite{Conformers} that the B2K-PLYP-D double-hybrid functional\cite{B2K-PLYP} with an empirical dispersion correction, in conjunction with a sufficiently large basis set, tracks the CCSD(T) reference data\cite{Conformers} for n-butane, n-pentane, and n-hexane exceedingly closely, and this is the approach we have followed for all systems with more than one conformer in this work.

Dispersion corrections for the DFT energies (denoted by the suffix "-D") were applied using our implementation of Grimme's expression\cite{Grimme1,Grimme2}.

\begin{equation}
E_{disp} = -s_{6} \sum_{i=1}^{N_{at}-1} \sum_{j=i+1}^{N_{at}} \frac{C_{6}^{ij}}{R_{ij}^{6}}f_{\rm dmp} \left(R_{ij} \right)\label{eq:grimme}
\end{equation}
where the damping function is taken as
\begin{equation}
 f_{\rm dmp} \left( R_{ij} \right) = \left[ 1 + \exp \left( -\alpha ( \frac{R_{ij}}{s_RR_{r}} -1 )\right)\right]^{-1}
\end{equation}
and $C_{6}^{ij}\approx\sqrt{C_{6}^{i}C_{6}^{j}}$, $R_{r}=R_{{\rm vdW},i}+R_{{\rm vdW},j}$  is the sum of the van der Waals radii of the two atoms in question, and the specific numerical values for the atomic Lennard-Jones constants $C_{6}^{i}$ and the van der Waals radii have been taken from Ref.\cite{Grimme1}, whereas the length scaling $s_R$=1.0 and hysteresis exponent $\alpha$=20.0 as per Ref.\cite{Grimme2}.

Eq.(\ref{eq:grimme}) has a single functional-dependent parameter, namely the prefactor $s_6$. This was taken from Refs.\cite{Grimme1,Grimme2} for BLYP, B3LYP, and PBE, from Ref.\cite{B2GP-PLYP} for the double hybrids, and optimized in the present work for the remaining functionals.  These were, for the most part, optimized against the S22 benchmark set of weakly interacting systems\cite{Hobza06}.

\section{Results and Discussion\label{sec:Res}}

\subsection{Overview of diagnostics for nondynamical correlation\label{subsec:Diagnostics}}
The percentages of the nonrelativistic, clamped-nuclei W1h total atomization energy at the bottom of the well (TAE$_e$) accounted for by SCF, and (T) triples contributions are reported in Table S1 of the Supporting Information, together with the coupled cluster ${\cal T}_1$ and ${\cal D}_1$ diagnostics\cite{Lee89,Lei00}, and the largest $T_2$ amplitudes. The percentage of the total atomization energy accounted for by parenthetical connected triple excitations, \%TAE$_e$[(T)], has been shown to be a reliable energy-based diagnostic for the importance of nondynamical correlation effects\cite{w4}. Ref.\cite{w4} gives useful criteria for assessing the extent of nondynamical correlation effects, e.g., \%TAE$_e$[SCF]$\ge$67\% and/or \%TAE$_e$[(T)]$<$2\% indicate systems that are dominated by dynamical correlation. 

As expected, all the alkanes considered in the present study exhibit very mild nondynamical correlation effects, and can be regarded as dominated by dynamical correlation. 77--79\% of the atomization energy is accounted for at the Hartree-Fock level, and 0.7--1.3\% by the (T) triples. Table S1 (see Supporting Information) shows that the \%TAE$_e$[(T)] slightly increases with the degree of branching. 

In systems with very mild nondynamical correlation, CCSD(T) is generally very close to the full CI limit, as the higher order triple excitations, $\hat{T}_3-$(T), and the connected quadruple excitations, $\hat{T}_4$, tend to largely cancel one another (they are of similar orders of magnitude, but connected quadruples universally increase atomization energy while higher-order triples generally decrease it\cite{w4,w4.4}). For the smaller systems (of up to five carbons), for which we were able to explicitly calculate the $\hat{T}_3-$(T) and (Q) contributions, the percentage of the atomization energy accounted for by post-CCSD(T) excitations, \%TAE[post-(T)], varies between -0.01 and -0.02\%. Small as these numbers may seem in relative terms, the atomization energy for an alkane with more than three carbons already exceeds 1,200 kcal/mol, 0.02\% of which amounts to 1 kJ/mol.

\subsection{Linear alkanes}
The gas phase heat of formation (viz. the atomization energy) of an arbitrary linear alkane can be obtained from the following hypohomodesmotic reaction involving ethane and propane.
\begin{equation}
{\rm C}_m{\rm H}_{2m+2}+(m-3){\rm C}_2{\rm H}_{6} \rightarrow  (m-2){\rm C}_3{\rm H}_{8}\label{reac:hypohomodesmotic}
\end{equation}
As the atomization energy of ethane is very well established both by W4 theory and by ATcT, it is highly desirable to also have a W4 reference value for propane. A component breakdown of the W4 atomization energies for methane, ethane, and propane is given in Table \ref{tab:W4comp}, while the final atomization energies at 0 K are compared with ATcT and CCCBDB experimental values in Table \ref{tab:W4comp}. There is good agreement between W4 and the available ATcT values (to within the sum of the uncertainties). 

W3.2lite component breakdowns from methane, ethane, propane, $n$-butane, and $n$-pentane (as well as for the relevant hypohomodesmotic reactions) are given in Table \ref{tab:W3.2lite}. Let us briefly consider the raw components of the W3.2lite atomization energies. First, we note that there is a perfect linear relationship (R$^2$$>$0.997) between all the components of the atomization energies and the number of carbons in the linear alkanes. The $\hat{T}_3-$(T) contribution reduces the atomization energies by amounts ranging form 0.12 kcal/mol in methane to 1.07 kcal/mol in $n$-pentane, and the (Q) contribution, which increases the atomization energy, ranges from 0.09 kcal/mol in methane to 0.79 kcal/mol in $n$-pentane. The overall post-CCSD(T) contributions, which increase linearly with the size of the system, reduce the atomization energies by 0.03--0.3 kcal/mol. We finally note that, while the directly computed W3.2lite total atomization energies in the hypothetical motionless state (``at the bottom of the well'') agree well with the available W4 data (to within 0.2 kcal/mol or better), agreement at 0 K is rather less pleasing (differences of 0.3, 0.6, and 0.7 kcal/mol are seen for methane, ethane, and propane, respectively). The chief part of these differences (0.2, 0.4, and 0.5 kcal/mol, respectively) comes from the comparatively low-level approximation used for the zero-point vibrational energies: in molecules like $n$-pentane, ZPVE reaches the 100 kcal/mol regime, and even a 1\% error due to neglect of explicit anharmonicity will translate into a 1 kcal/mol error in the final ZPVE. We made the point earlier\cite{C2H6} that, at least for species containing many hydrogens, the factor limiting accuracy of W$n$ and similar thermochemical protocols will increasingly be the quality of the zero-point vibrational energy. Nevertheless, for the `quasi-W4' data obtained in the present paper from  hypohomodesmotic or isodesmic cycles, the ZPVEs that enter the final results are the very accurately known ZPVEs of methane, ethane, and propane on the one hand, and the small ZPVE component (nearly two orders of magnitude smaller than for the brute-force TAE calculation) of the hypohomodesmotic or isodesmic reaction energy on the other hand. The relatively low-level approximation to the latter simply cannot wreak as much "damage" in the latter case.

Turning to the hypohomodesmotic reactions of $n$-butane and $n$-pentane, the most striking feature of Table \ref{tab:W3.2lite} is the near-perfect cancellation of {\em all} the valence correlation contributions between reactants and products. For $n$-butane, the CCSD, (T), $T-$(T), and (Q) contributions to the hypohomodesmotic reaction energy amount to merely, -0.03, -0.03, -0.01, and 0.00 kcal/mol, respectively. While for the hypohomodesmotic reaction involving $n$-pentane they amount to: +0.02, -0.02, -0.03, and +0.01 kcal/mol, respectively. The scalar relativistic and DBOC contributions are, likewise,  basically null. The dominant contributions to the reaction energies come from the SCF, inner-shell, and ZPVE components (specifically, +0.20, +0.07, and +0.13 kcal/mol, respectively, for the $n$-butane reaction, and +0.21, +0.11, and +0.22 kcal/mol, respectively, for the $n$-pentane reaction). Overall, these hypohomodesmotic reactions are very slightly endothermic; the reaction energies at 0 K are 0.32 and 0.52 kcal/mol for $n$-butane and $n$-pentane, respectively. 

Having established that valence post-CCSD(T) and DBOC contributions to the hypohomodesmotic reaction energies are thermochemically negligible, we proceed to calculate the atomization energies of larger $n$-alkanes from W1h hypohomodesmotic reaction energies. W1h component breakdowns from linear alkanes up to $n$-octane (as well as for the relevant hypohomodesmotic reactions) are gathered in Table S2 of the Supporting Information. First, let us consider the components of the atomization energies for the five species for which a comparison with W3.2lite can be made. The SCF component, extrapolated from the cc-pV\{T,Q\}Z basis sets, is well converged. The valence CCSD contribution, extrapolated from the same basis sets, systematically overestimates the cc-pV\{Q,5\}Z results (by 0.4, 0.5, 0.7, 0.9, and 1.1 kcal/mol for methane, ethane, propane, $n$-butane, and $n$-pentane, respectively). These differences are partially compensated by the fact that the W1h inner-shell contribution calculated with the MTsmall basis set systematically underestimates the cc-pwCV\{T,Q\}Z results (by $\sim$0.1 kcal/mol for methane and ethane, and by $\sim$0.2 kcal/mol for the larger alkanes). The valence (T) component, extrapolated from the cc-pV\{D,T\}Z basis sets, is reasonably converged---the largest deviation (of 0.1 kcal/mol) from the cc-pV\{T,Q\}Z results is seen for neopentane. Overall, the W1h atomization energies at 0 K overestimate the W3.2lite values by 0.2, 0.3, 0.4, 0.6, and 0.7 kcal/mol for methane, ethane, propane, $n$-butane, and $n$-pentane, respectively. This demonstrates the limitations of directly computing TAEs with lower level compound thermochemistry methods such as W1h without the aid of hypohomodesmotic or isodesmic cycles.
(We note that most of the difference comes from basis set incompleteness in the W1h valence CCSD contribution, and that its nefarious effects is actually mitigated by error compensation with neglect of DBOC and post-CCSD(T) components. By way of illustration, the W2.2h numbers for n-butane and n-pentane are 1220.71 and 1497.75 kcal/mol, respectively, very close to W3.2lite.)

Turning to the W1h components of the hypohomodesmotic reaction energies (Table S2, Supporting Information) the SCF, valence CCSD, valence (T), and inner-shell contributions to the hypohomodesmotic reaction energies are very similar to their counterparts at the W3.2lite level. The overall reaction energies at 0 K differ by merely 0.02 and 0.03 kcal/mol for the $n$-butane and $n$-pentane hypohomodesmotic reactions, indicating that the basis sets used in W1h are sufficiently large to ensure adequate convergence  of these hypohomodesmotic reaction energies. For the larger $n$-alkanes, again we see that the valence (T) contribution practically cancels out between reactants and products. The valence CCSD contribution to the reaction energy becomes thermochemically significant for the larger $n$-alkanes, reaching 0.30 kcal/mol for the $n$-octane reaction. The overall reaction energies at 0 K range from 0.34 to 1.50 kcal/mol for the $n$-butane and $n$-octane reactions, respectively. 

Using W4 atomization energies for ethane and propane and assuming that the hypohomodesmotic reaction energies stay constant between W1h and W4 theories, we obtain `quasi-W4' atomization energies. Table \ref{tab:TAEhypo} compares the `quasi-W4' atomization energies obtained from W3.2lite and W1h reaction energies with experimental data taken from CCCBDB\cite{cccbdb} (adjusted for the revised, ATcT, heat of formation of carbon atom). First of all, for the species where such a comparison can be made, the `quasi-W4' values obtained from W3.2lite and W1h reaction energies are in very close agreement with each other. The hypohomodesmotic `quasi-W4' and adjusted CCCBDB values agree very well (0.06--0.32 kcal/mol of one another, where the largest deviations are seen for $n$-pentane).

Finally, we note that the W1h TAE$_0$ increasingly overestimate the `quasi-W4' TAE$_0$ (by 1.3--2.0 kcal/mol) on going from $n$-butane to $n$-octane, and even the W3.2lite TAE$_0$ for $n$-butane and $n$-pentane overestimate the `quasi-W4' TAE$_0$ by 0.7 and 0.8 kcal/mol, respectively. These large differences demonstrate the obvious advantage of using hypohomosesmotic reactions rather than atomization reactions when one is limited to comparatively low-level compound thermochemistry methods such as W1h. The difference between the W3.2lite and best values mostly derives from neglect of explicit anharmonicity in the ZPVE. (The use of scale factors, of course, to some degree accounts implicitly for anharmonicity.)

\subsection{Branched alkanes}
In the nature of things, for an arbitrary alkane, there are more isodesmic reactions involving small hydrocarbon prototypes then there are hypohomodesmotic ones, rendering the former more useful for thermochemical applications. If we consider only prototypes for which we have explicit W4 reference values (namely, methane, ethane, and propane) then for an arbitrary alkane two linearly-independent isodesmic reactions are: reaction \ref{reac:hypohomodesmotic} and the bond-separation reaction \ref{reac:isodesmic}, 
\begin{equation}
{\rm C}_m{\rm H}_{2m+2}+(m-2){\rm C}{\rm H}_{4}   \rightarrow  (m-1){\rm C}_2{\rm H}_{6}\label{reac:isodesmic}
\end{equation}
Table \ref{tab:W3.2lite} gives W3.2lite component breakdowns from isobutane and neopentane (as well as for the said isodesmic reaction energies). We note that for both isodesmic reactions the post-CCSD(T) contributions are not only fairly small, but partly cancel out between higher-order triples and connected quadruples, something more pronounced for reaction \ref{reac:hypohomodesmotic} than for reaction \ref{reac:isodesmic}. 

Table S2 of the Supporting Information gives W1h total atomization energies component breakdowns from branched alkanes of up to eight carbons and Table S3 of the Supporting Information gives W1h component breakdowns from the said isodesmic reaction energies. As was the case for the hypohomodesmotic reactions, the W1h components of reaction \ref{reac:hypohomodesmotic} are practically identical to their counterparts at the W3.2lite level, indicating that the basis sets used in W1h theory are large enough to ensure convergence of the components of this isodemic reaction. The components of the bond-separation reaction \ref{reac:isodesmic} are not as similar to their counterparts at the W3.2lite level---the largest deviations are seen for the CCSD component (namely, 0.14 and 0.20 kcal/mol for the reactions involving isobutane and neopentane). Again we find that the DBOC contributions, which are neglected in W1h, are fairly small, something more pronounced for reaction \ref{reac:hypohomodesmotic} than for reaction \ref{reac:isodesmic}. 

Perusing the energy components of reactions \ref{reac:hypohomodesmotic} and \ref{reac:isodesmic} in Table S3 (see Supporting Information), several systematic features emerge: (a) the valence CCSD and (T) components of reaction \ref{reac:hypohomodesmotic} are substantially lower than those of reaction \ref{reac:isodesmic}; (b) inner-shell correlation effects are somewhat more pronounced in reaction \ref{reac:hypohomodesmotic} than in reaction \ref{reac:isodesmic}; (c) scalar relativistic and DBOC contributions are fairly small, but in contrast to reaction \ref{reac:hypohomodesmotic}, in reaction \ref{reac:isodesmic} they systematically increase with the size of the alkane; and (e) the ZPVE contribution to reaction \ref{reac:hypohomodesmotic} is substantially lower than to reaction \ref{reac:isodesmic}. 

In effect, the energy components (except for inner-shell) of reaction \ref{reac:hypohomodesmotic} cancel out to a larger extent between reactants and products than in the bond-separation reaction \ref{reac:isodesmic}. Overall, the reaction energies at 0 K for reaction \ref{reac:hypohomodesmotic} are considerably lower (by 6--17 kcal/mol) than those of reaction \ref{reac:isodesmic}. The superiority of reaction \ref{reac:hypohomodesmotic} is ascribed to the fact that the numbers of CH$_2$ and CH$_3$ groups are roughly equal on both sides of the reaction (with the notable exception of hexamethylethane, a.k.a. 2,2,3,3-tetramethylbutane), while in reaction \ref{reac:isodesmic} the difference in CH$_2$ and CH$_3$ groups between reactants and products increases systematically with the size of the alkane. Furthermore, in reaction \ref{reac:hypohomodesmotic} the number of 1,3-interactions is roughly equal on both sides, while in reaction \ref{reac:isodesmic} 1,3-interactions occur only on the left-hand-side, and thus the imbalance increases systematically with the size of the alkane. The difference between the two reactions is further emphasized when considering the linear alkanes for which reaction \ref{reac:hypohomodesmotic} is hypohomodesmotic, and thus the numbers of CH$_2$ groups,  CH$_3$ groups, and 1,3-interactions are perfectly balanced on both sides of the reaction. For example, for $n$-octane, the SCF, CCSD, and (T) components of reaction \ref{reac:isodesmic} are 4.79, 6.87, and 1.56 kcal/mol, while for reaction \ref{reac:hypohomodesmotic} they are merely 0.27, 0.30, and 0.03 kcal/mol, respectively (see Table S2 of the Supporting Information).
 
From the linearly independent reactions \ref{reac:hypohomodesmotic} and \ref{reac:isodesmic} we can easily construct new isodesmic reactions---any linear combination of isodesmic reactions is also isodesmic---in which the number of 1,3-interactions, CH$_2$ groups, or CH$_3$ groups are perfectly balanced on both sides of the reaction. In the remainder of the paper, these are referred to as (1,3), (CH$_2$), and (CH$_3$) reactions, respectively. The question that naturally arises is: which of the five isodemic reactions should be used to derive the `quasi-W4' atomization energy of the branched alkanes? As far as the electronic structure is concerned, we can divide the question into two parts: convergence of the n-particle space ("correlation treatment") and one-particle space ("basis set"). (In the present study we are not analyzing errors arising from anharmonic corrections to the ZPVE.) 

We have already seen for isobutane and neopentane (Table \ref{tab:W3.2lite}) that convergence of the n-particle space is faster in reaction \ref{reac:hypohomodesmotic} than in reaction \ref{reac:isodesmic}. Table \ref{tab:W3.2lite} also shows that the isodesmic reaction that balances the 1,3-interactions exhibits the fastest convergence, i.e., the $\hat{T_3}-$(T) and (Q) contributions are 0.00 and 0.00 kcal/mol for the isobutane reaction, and -0.01 and -0.02 kcal/mol for the neopentane reaction. 

Convergence of the one-particle space can be explored in a more systematic manner. Table S4 (given in the Supporting Information) shows the basis set convergence of the SCF, CCSD, and (T) components of the said isodesmic reactions. In general, the SCF component converges quite rapidly with the basis set size, for all the isodesmic reactions the cc-pV\{D,T\}Z results are close to the basis set limit, something more pronounced for reactions \ref{reac:hypohomodesmotic} and (1,3) (we note that reaction \ref{reac:isodesmic} exhibits anomalous convergence for some of the systems). Convergence of the valence CCSD component becomes markedly slower. For most of the systems the isodesmic  reactions \ref{reac:hypohomodesmotic} and (1,3) the cc-pVTZ basis set yields results close to the basis set limit; nevertheless, they exhibit anomalous, nonmonotonous, basis set convergence, the cc-pV\{D,T\}Z results being further away from the basis set limit. As for the valence (T) component, again, these two reactions converge more rapidly, i.e., for most systems convergence is obtained even with the cc-pVDZ basis set. 

Table S3 of the Supporting Information gives W1h component breakdowns from the five isodesmic reaction energies. Several systematic trends are observed: (a) for any branched alkane the reaction energies both in the hypothetical motionless state ("at the bottom of the well") and at 0 K increase in the following order: RE[(1,3)] $<$ RE[\ref{reac:hypohomodesmotic}] $<$ RE[(CH$_3$)] $<$ RE[(CH$_2$)] $<$ RE[\ref{reac:isodesmic}]; (b) the SCF, valence CCSD, valence (T), relativistic, and DBOC reaction components, generally, increase in the same order; (c) the inner-shell and ZPVE corrections decrease in the same order. 

From the above discussion, it seems that reactions \ref{reac:hypohomodesmotic} and (1,3) offer the greatest similarity between the electronic structure of the reactants and products. The former has the additional advantage that it results in atomization energies with smaller uncertainties (Table S3, Supporting Information). Therefore, we select it as the `best' reaction for deriving the most accurate `quasi-W4' atomization energies. 

Table S3 (see Supporting Information) gives the TAE$_0$ of the branched alkanes obtained from the five said isodesmic reactions (by assuming that the isodesmic reaction energy stay constant at the W1h and W4 levels, and using the W4 TAE$_0$ for methane, ethane and propane). In practice, the resulting TAE$_0$ vary by 0.3--3.2 kcal/mol depending on which reaction is used and the standard deviation varies between 0.2 and 1.1 kcal/mol. It is interesting to note, however, that for any alkane the TAE$_0$ are ordered in the same way as the reaction energies, namely, TAE[\ref{reac:isodesmic}] $<$ TAE[(CH$_2$)] $<$ TAE[(CH$_3$)] $<$ TAE[\ref{reac:hypohomodesmotic}] $<$ TAE[(1,3)]. 

Table \ref{tab:TAEhypo} compares the final `quasi-W4' atomization energies obtained from isodesmic reaction \ref{reac:hypohomodesmotic} with experimental data taken from the CCCBDB\cite{cccbdb}: these were adjusted for the revised (ATcT) heat of formation of carbon atom\cite{w4}. First, the `quasi-W4' TAE$_0$ obtained from W3.2lite and W1h reaction energies are in close agreement with each other. The isodesmic `quasi-W4' and adjusted CCCBDB values agree very well (to within 0.1--0.3 kcal/mol) for all the systems but neopentane. The discrepancy of 1.1  kcal/mol seen for neopentane is too large to be easily explainable in terms of issues with the calculations. We note, however, that a discrepancy of 0.7 kcal/mol exists between our best calculated H$_{298}-$H$_0$ and the value that Scott\cite{BOM666} used in TRC and propagated into CCCBDB; moreover, the computed and observed $\Delta H^\circ_{f,298K}$ are in much more plausible agreement. (We also note that neopentane does not have conformers and therefore these cannot introduce uncertainty in H$_{298}-$H$_0$. In addition, the Ayala-Schlegel\cite{AyalaSchlegel} internal rotation correction is an order of magnitude smaller than the discrepancy.) We would argue that the experimental data for neopentane bear reexamination. 

Table \ref{tab:TAEhypo} also compares the final `quasi-W4' heats of formation at 298 K with the available experimental data. There is reasonable agreement between theory and experiment (generally, to within 0.0--0.4 kcal/mol). In general, the available TRC values are 0.1--0.3 kcal/mol higher than the `quasi-W4' values, and the Rossini values are 0.0--0.4 kcal/mol higher than the theoretical values. 

The equilibrium geometry of isooctane (a.k.a. 2,2,4-trimethylpentane, the ``100\%'' fixpoint of the ``octane rating'' scale) has no symmetry: for this reason we were only able to obtain a W1h value for the first-order saddle point (which has $C_s$ symmetry) possessing an imaginary frequency (37.5i cm$^{-1}$) that corresponds to an internal rotation. The W1h TAE$_e$ for the idealized structure (Table S2, Supporting Information) is 2483.28 kcal/mol. For the deformation energy difference between $C_1$ isooctane and the $C_s$ saddle point we obtain 0.25 kcal/mol at the CCSD(T) limit (0.135, 0.100, and 0.015 kcal/mol from the SCF, valence CCSD, and valence (T) components, respectively.) Assuming that the core-valence and relativistic contributions to the deformation energy will be zero, we obtain an estimated W1h TAE$_e$ of 2483.53 kcal/mol for the equilibrium structure of isooctane. Inclusion of the ZPVE from a scaled B3LYP/pc2 harmonic calculation (150.69 kcal/mol) results in a W1h TAE$_0$ of 2332.84 kcal/mol. Using the reaction energy of reaction \ref{reac:hypohomodesmotic} at the W1h level, and W4 atomization energies for ethane and propane, we obtain a `quasi-W4' atomization energy for isooctane of 2330.67 kcal/mol, in reasonable agreement with the NIST value of 2330.94 kcal/mol (adjusted for the revised, ATcT, heat of formation of carbon atom).

\subsection{Performance of compound thermochemistry methods for alkanes}
Table \ref{tab:Gn} presents root mean square deviations (RMSD) for atomization energies relative to our best values for more approximate compound thermochemistry methods such as G2(MP2)\cite{g2mp2}, G2\cite{g2}, G3\cite{g3}, G3B3\cite{g3b3}, G4\cite{g4}, G4(MP2)\cite{g4mp2}, CBS-QB3\cite{cbs-qb3}, CBS-APNO\cite{cbs-apno}, W1h\cite{w1h}, W2.2h, and W3.2lite. Application of the more expensive W2.2h and W3.2lite methods was possible only for a subset of small systems. Starting with the zero-point exclusive (`bottom of the well') data, the performance of the empirically corrected G$n$ methods systematically improves as one proceeds along the series (the RMSDs are: G1 11.2, G2 4.3, G3 3.7, and G4 1.3 kcal/mol). G2(MP2) performs significantly worse (RMSD$=$5.9) than the standard G2 procedure; interestingly, standard G4 offers no improvement over G4(MP2). We note that, while G2(MP2) does not include post-MP2 correlation effects at all, the somewhat confusingly named G4(MP2) does include a small basis set CCSD(T) step: the main differences with full G4 theory are the absence (vs. presence) of an explicit inner-shell correlation step and of valence MP4 steps with some intermediate-sized basis sets. As in the n-alkanes the inner-shell correlation term scales quite linearly with $n$, this is an optimal scenario for absorbing its effect into the empirical `high-level correction', while the systems are also sufficiently dominated by dynamical correlation (as well as apolar) that the MP$n$ series converges well and a single CCSD(T)/6-31G* step can adequately handle post-MP2 correlation effects.

W1h gives a RMSD of 0.9 kcal/mol for the whole set and 0.7 kcal/mol for the subset of small systems. Using more elaborate basis sets for the extrapolations of the SCF, CCSD, and (T) contributions in W2.2h cuts the RMSD for the smaller subset to 0.3 kcal/mol. Including post-CCSD(T) correlation effects in W3.2lite further reduces the RMSD to 0.2 kcal/mol. For the empirical methods the RMSD for the isomerization energies are lower than for the atomization energies: all the methods show similar performance with RMSD of 0.7--1.2 kcal/mol. 
Interestingly, while G4 outperforms both CBS-QB3 and CBS-APNO for the `bottom of the well' atomization energies, CBS-QB3 and CBS-APNO both surpass G4's performance for the isomerization energies.

We have already stressed that for molecules containing many hydrogens the principal factor limiting the accuracy of W$n$ methods lies in the quality of the zero-point vibrational energy. This is in accord with the deterioration in performance of the W$n$ methods when zero-point corrections are included, the RMSD increase by 0.6 kcal/mol for  W1h, W2.2h, and W3.2lite compared to the zero-point exclusive RMSD. Despite that both G$n$ and W$n$ methods use ZPVEs at relatively low levels of theory (scaled HF/6-31G* and B3LYP/pc-2 harmonic frequencies, respectively) the performance of the G$n$ methods significantly improves due to the fact that these methods were parametrized against experimental atomization energies at 0 K. Thus, some correction for the zero-point energy is evidently absorbed in the empirical corrections. What's more, the systematic convergence of the G$n$ methods seen for the zero-point exclusive results is not observed for the zero-point inclusive results: G3 shows the best performance with a RMSD of 0.5 kcal/mol, while G2 and G4 give RMSDs of 0.8 and 0.7 kcal/mol, respectively. Again we see that G4 and G4(MP2) show similar performance and that G2(MP2) performs substantially worse than G2.

\subsection{Performance of density functional theory for alkanes}
Recent studies\cite{Yao03,Gilbert05,Radom05,Grimme06,Schle06,Schre07} have shown that DFT methods fail to adequately predict dissociation\cite{Yao03,Gilbert05,Radom05}, isomerization\cite{Grimme06}, and isodesmic\cite{Schle06} reaction energies involving alkanes. Grimme\cite{Grimme06} showed that DFT methods generally predict the wrong sign for the $n$-octane$\rightarrow$hexamethylethane isomerization, with errors ranging from 5.4 (B2-PLYP) to 11.8 (BLYP). Schleyer and coworkers\cite{Schle06} investigated the performance of various DFT functionals in reproducing the experimental reaction energy of the isodesmic reaction \ref{reac:isodesmic} for linear alkanes of up to $n$-decane. They showed that van der Waals corrected DFT functionals (such as MPWB1K and MPW1B95) underestimate the protobranching stabilization energy as defined by reaction \ref{reac:isodesmic} by $\sim$1 kcal/mol, and other functionals by up to $\sim$2 kcal/mol, per protobranch. 

In the present section we compare the relative performance of different DFT exchange-correlation functionals in predicting atomization, isomerization, and isodesmic reaction energies. The XC functionals employed include the following classes (numbered by `rung' on the Perdew ladder\cite{JacobsLadder}): (1) the local density approximation (LDA), specifically, SVWN5\cite{SVWN5}; (2) the pure generalized gradient approximation (GGA) functionals: BLYP\cite{B88ex,LYPc}, PBE\cite{PBE}, and HCTH407\cite{HCTH407}; (3) the meta GGAs: TPSS\cite{TPSS} and M06-L\cite{M06-L}; (4a) the hybrid GGAs (which one might term `imperfect fourth-rung functionals'): B3PW91\cite{B3,PW91c}, B3LYP\cite{LYPc,B3,B3LYP}, B97-1\cite{B97-1}, PBE0\cite{PBE0}, B97-2\cite{B97-2}, B97-3\cite{B97-3}, and X3LYP\cite{X3LYP}; (4b) the hybrid meta-GGAs: B1B95\cite{B1B95}, TPSSh\cite{TPSSh}, BMK\cite{BMK}, TPSS1KCIS\cite{TPSS1KCIS}, PW6B95\cite{PW6B95}, M06\cite{M06}, M06-2X\cite{M06}; and (5) the double hybrid functionals: B2-PLYP\cite{B2-PLYP}, mPW2-PLYP\cite{mPW2-PLYP}, B2T-PLYP\cite{B2K-PLYP}, B2K-PLYP\cite{B2K-PLYP}, and B2GP-PLYP\cite{B2GP-PLYP}. For the conventional functionals we use the pc-2 basis set of Jensen\cite{pc-2}, which is of $[4s3p2d1f]$ quality but optimized for Hartree-Fock and DFT, while for MP2 and the double hybrids which exhibit slower basis set convergence we also use the pc-3 basis set. 

To ensure we are ``comparing apples to apples'', so to speak, secondary effects that are not explicitly included in the DFT calculations such as relativity, deviations from the Born-Oppenheimer approximation, and zero-point vibrational corrections are excluded from the reference values. For methane, ethane, and propane we use nonrelativistic, clamped-nuclei, zero-point exclusive TAEs from W4 theory, and for the remaining systems our best `quasi-W4' values (given in the second or third column of Table \ref{tab:TAEhypo}). The B3LYP/pc-2 reference geometries, the said reference TAEs, and the individual errors of the various functionals can be found in the Supporting Information. The empirical $s_6$ scaling factors for each functional were taken from ref. \cite{B2GP-PLYP,Grimme1,Grimme2} or otherwise for SVWN5, PBE, HCTH407, BLYP, TPSS, B97-1, B97-2, B97-3, TPSSh, TPSS1KCIS, PW6B95, and B1B95 optimized using the same procedure as detailed in Ref. \cite{B2GP-PLYP}. The optimized $s_6$ values, as well as error statistics over the S22 benchmark set of weakly interacting systems by Hobza and coworkers\cite{Hobza06}, can be found in Table S5 of the Supporting Information. The RMSD, mean signed deviations (MSD), and mean average deviations (MAD) for the atomization reactions are gathered in Table \ref{tab:DFTatomization}. Table \ref{tab:DFTisomerization} lists the RMSD, MSD, and MAD for the $n$-alkane$\rightarrow$branched-alkane isomerization reactions with and without dispersion corrections. Finally, we consider the performance of DFT for the isodesmic reaction \ref{reac:isodesmic}. Table \ref{tab:DFTisodesmic} reports the RMSD for the $n$-, iso-, and neo-alkanes as well as for the entire set. 

Before proceeding to a detailed discussion of the specific performance of the various DFT functionals for our four evaluation sets, a few general remarks are in order:
\begin{itemize}
\setlength{\itemsep}{0pt}
\item The S22 benchmark set for weak interactions contains 22 model complexes involving typical noncovalent interactions, such as hydrogen bonding (e.g., H$_2$O and NH$_3$ dimers), dipolar and multipolar dispersion interactions (e.g., pyrazine and benzene dimers). We have previously\cite{B2GP-PLYP} made the point that small $s_6$ values can be seen as an indication of a functional's ability to cope with dispersion. This is demonstrated by an almost perfect linear correlation between the magnitude of the $s_6$ values and the uncorrected RMSD over the S22 set seen in Table S5 (see Supporting Information). Excluding the M06 family of functionals (M06, M06-L, and M06-2X) and the double hybrids (B2K-PLYP, B2GP-PLYP, B2-PLYP, and mPW2-PLYP) all the functionals perform rather poorly without the dispersion correction: RMSD vary between 3--6 kcal/mol. Correcting for dispersion dramatically reduces the RMSD to 0.3--1.0 kcal/mol. For comparison, Hartree-Fock gives a RMSD of 0.8 kcal/mol after correcting for dispersion. 
\item Of our four evaluation sets, the atomization energies are clearly the toughest nut to crack. In general, {\em a posteriori} correction for dispersion are essential for an accurate estimation of the atomization energies. As expected, the dispersion corrections increase with the number of carbons in the alkane, and for isomers with the number of 1,3-interactions present. Nevertheless, the magnitude of the corrections are somewhat surprising: for instance, with $s_6$=1.0, the dispersion energy correction ranges from 0.6 (methane) to 27.0(!) kcal/mol (hexamethylethane). The latter amounts to 1\% of the total atomization energy. 
\item Without the dispersion correction the RMSD for the $n$-alkane$\rightarrow$branched-alkane isomerization reactions (Table \ref{tab:DFTisomerization}) ranges from 0.2 kcal/mol (M06-2X) to 5.4 kcal/mol (HCTH407). Inspection of the individual errors reveals that most functionals predict the wrong sign for the isomerization reactions of the larger alkanes: $n$-hexane, $n$-heptane, and $n$-octane (i.e., that the linear isomer is more stable than the branched isomer). MP2, SVWN5 (!), and the M06 family of functionals (M06, M06-L, and M06-2X) are the only ones to obtain the right sign across the board. The corrected results are much more encouraging: without exception, all the functionals predict that the branched isomers are more stable than their linear counterparts, and for most of the functionals the overall RMSD is below 0.5 kcal/mol, specifically X3LYP, B97-3, M06-2X, B2K-PLYP, and B2GP-PLYP with a RMSD below 1 kJ/mol.
\item As for the performance of DFT for the isodesmic reaction \ref{reac:isodesmic} (Table \ref{tab:DFTisodesmic}), we find that all functionals other than LDA underestimate the reaction energy (whether it involves linear or branched alkanes), confirming the findings of Schleyer and coworkers\cite{Schle06} for linear alkanes. Also we find that without correction for dispersion the RMSD increases with the number of 1,3-interactions present, i.e., in the order $n$-alkanes$<$isoalkanes$<$neoalkanes. This trend is of course attributed to the increase in dispersion interactions with the degree of branching. 
\end{itemize}

We shall start with the nonempirical functionals SVWN5 LDA, PBE GGA, and TPSS meta-GGA functionals and their corresponding hybrid functionals PBE0 and TPSSh. The SVWN5 functional fails miserably for the alkane atomization energies, with a RMSD of over 200 kcal/mol, and will be omitted from further discussion, other than to say that it performs surprisingly well for the isomerization and isodesmic reactions. For the S22 set the nonempirical functionals perform rather poorly with uncorrected RMSD between 4--5 kcal/mol and corrected RMSD between 0.7--1.0 kcal/mol. For the atomization reactions (Table \ref{tab:DFTatomization}) the nonempirical functionals TPSS, PBE, TPSSh, and to a lesser extent PBE0 systematically overbind the alkanes (and in fact, are the only functionals other than LDA that do so) as evident from MSD$\approx$MAD. Obviously, there is no point in `correcting' for dispersion then, as it can only increase the errors further. Indeed, if the empirical $s_6$ scaling factors are reoptimized by minimizing the RMSD for the atomization energies, then negative (anomalous) $s_6$ values are obtained. The uncorrected RMSD for the atomization reactions are 4.2, 13.7, 3.7, and 1.7 kcal/mol for TPSS, PBE, TPSSh, and PBE0, respectively (we note that without the dispersion correction PBE0 exhibits the best performance of the functionals considered). For the linear$\rightarrow$branched isomerization reactions the uncorrected RMSD are on the order of 3 kcal/mol and the corrected RMSD vary between 0.3--0.9 kcal/mol, where PBE0 and PBE are the best performers. For the isodesmic reaction \ref{reac:isodesmic}, the uncorrected RMSD vary between 5--7 kcal/mol and the corrected RMSD between 0.4--1.3 kcal/mol, again with PBE and PBE0 as best performers. 

In the following discussion the empirical functionals are conveniently divided into three categories: lightly, moderately, and heavily parameterized. The GGA BLYP, hybrid GGA B3PW91 and B3LYP, and hybrid meta GGA TPSS1KCIS and B1B95 functionals belong to the first category. The uncorrected RMSD over our four validation sets (S22, atomization, isomerization, and isodesmic reactions) are unacceptably large ranging from 2--37 kcal/mol. After applying the dispersion correction, B3LYP emerges as the best performer with RMSD of 0.7, 1.9, 0.5, and 0.6 kcal/mol for the said four validations sets, respectively, followed by B1B95 with RMSD of 0.5, 2.5, 0.8, and 0.9 kcal/mol, respectively. We note that BLYP, B3PW91 and TPSS1KCIS yield comparable RMSDs for the S22, isomerization, and isodesmic reactions, but  the former grossly underestimates the atomization energies and the latter two largely overestimate the atomization energies. 

The X3LYP and PW6B95 functionals may be regarded as belonging to the second category. Again the RMSD before correcting for dispersion are unacceptably large (2--13 kcal/mol). After correcting for dispersion both functionals perform relatively well, particularly PW6B95 with RMSD of 0.5, 0.2, 0.3, and 1.6 kcal/mol for the S22, atomization, isomerization, and isodesmic reactions, respectively. 

The heavily parametrized functionals include the GGA HCTH407, meta GGA M06-L, hybrid GGAs B97-1, B97-2, B97-3, and hybrid meta GGAs BMK, M06, and M06-2X. The M06 family of functionals (M06-L, M06, and M06-2X) have exceptionally low $s_6$ values (0.20, 0.25, and 0.06, respectively).  Without the dispersion correction M06-2X performs relatively well for the S22 and isomerization reactions, but grossly underestimates the atomization energies. All the other functionals yield unacceptably large errors without the dispersion correction. After correcting for dispersion, BMK and B97-2 seem to offer the best performance with BMK RMSDs of \{0.6, 1.9, 1.3, 0.6\} kcal/mol and B97-2 RMSDs of \{0.5, 2.6, 0.4, 0.6\} kcal/mol
for the S22, atomization, isomerization, and isodesmic datasets.

Finally, we come to the double-hybrid class of functionals (which may be considered lightly parameterized). It is perhaps not surprising that (for the BLYP-based double hybrids) the $s_6$ values decrease with the percentage of MP2 correlation included. Furthermore, the performance of the double hybrides (over the four evaluation test sets) systematically improves with the percentage of MP2 correlation included: most notably for the atomization energies RMSD of 12.5, 8.7, 8.0, 5.5, and 4.2 kcal/mol are obtained with the mPW2-PLYP (25\% MP2), B2-PLYP (27\% MP2), B2T-PLYP (31\% MP2), B2GP-PLYP (36\% MP2), and B2K-PLYP (42\% MP2) functionals, respectively. 
(For this narrow application, the original `thermochemistry and H-transfer barriers' parametrized double hybrid B2K-PLYP\cite{B2K-PLYP} thus slightly outperforms the more `general-purpose' parametrized B2GP-PLYP\cite{B2GP-PLYP}.)
Similar trends are  seen for the other three evaluation sets. 
After correction for dispersion, the performance of the double hybrids is quite remarkable: 
for the four validation sets (S22, atomization, isomerization, and isodesmic reactions),
B2GP-BLYP yields RMSD of 0.4, 0.2, 0.2, and 0.5 kcal/mol, respectively, while B2K-PLYP slightly bests it for the isomerization energies and slightly underperforms it for the isodesmic reactions (0.1 kcal/mol). 
We note that B2T-BLYP and B2-PLYP yield similar RMSD for the S22, isomerization, and isodesmic reactions, but an RMSD of $\sim$1.0 kcal/mol for the atomization reactions. 

As a numerical experiment, we reoptimized the empirical $s_6$ scaling factor to minimize the RMSD for the atomization reactions (Table \ref{tab:DFTatomization}). Obviously this approach also corrects for deficiencies other than dispersion, such as basis set incompleteness and/or limitations of the XC functional. Nevertheless, for PW6B95, B2GP-PLYP, and B2K-PLYP there was no significant change in the $s_6$ values or RMSD, indicating that (a) the original $s_6$ values are optimal; and (b) that these functionals do not exhibit any systematic errors. We note that for a few other functionals, namely  X3LYP, B3LYP, BMK, B2T-PLYP, and B2-PLYP, the reoptimized $s_6$ values are very similar to the original ones but the RMSD is improved by 0.6--1.2 kcal/mol.

\subsection{Hydrocarbon prototype reactions}
As discussed by Allen and coworkers\cite{Schleyer09}, the heat of formation at 0 K (viz. the atomization energy) of an arbitrary acyclic hydrocarbon can be obtained from hypohomodesmotic reactions involving a very limited number of branching prototypes: methane, ethane, ethene, acetylene, propane, propene, propyne, allene, isobutane, neopentane, and isobutene. As the atomization energies of the first four species are very well established both by W4 theory and by ATcT, this leaves seven species for which it would be highly desirable to have highly accurate reference values. W4 component breakdowns from methane, ethane, ethene, acetylene, propane, propene, propyne, and allene are given in Table \ref{tab:W4comp}. The final W4 atomization energies (Table \ref{tab:W4comp}) are: 392.47, 666.18, 532.09, 388.70, 942.95, 811.53, 670.45, and 669.37 kcal/mol, respectively. For isobutane, neopentane, and isobutene we obtain `quasi-W4' TAE$_0$ from isodesmic reaction energies at the W3.2level by using reaction \ref{reac:hypohomodesmotic} for the former two and isobutene+CH$_4$$\rightarrow$C$_2$H$_4$+C$_3$H$_8$ for the latter (component breakdowns from the said reactions are given in Table \ref{tab:W3.2lite}). Using W4 reference values for ethane, propane, and ethene we obtain atomization energies of 1221.73, 1501.01, and 1092.10 kcal/mol for isobutane, neopentane, and isobutene, respectively.

\section{Conclusions}
Benchmark post-CCSD(T) atomization energies and heats of formation for C$_n$H$_{2n+2}$ (all isomers up to $n=$6 inclusive, plus selected isomers for $n=$7 and $n=$8) have been obtained. Excellent agreement with the best available experimental data has been observed except that an issue with the experimental enthalpy function of neopentane has been identified. 

The hypohomodesmotic reaction energy (eq. \ref{reac:hypohomodesmotic}, in conjunction with linear alkanes) converges very rapidly with the level of theory. Valence post-CCSD correlation effects are thermochemically negligible (at least for the small alkanes considered in the present work). Basis set requirements are likewise rather modest the SCF component converges with the cc-pV\{D,T\}Z basis sets and CCSD component with the cc-pV\{T,Q\}Z basis sets. Scalar relativistic and DBOC contributions are thermochemically insignificant. Inner-shell correlation accounts for 25--30\% of the reaction energy at the `bottom of the well'. 

For isodesmic reactions convergence of the one- and n-particle spaces depends on the nature of the reaction: it is found that balancing the number of 1,3-ineractions between reactants and products significantly accelerates the convergence. 

We evaluated the performance of popular compound thermochemistry methods in reproducing the atomization energies. Post-CCSD(T) corrections are necessary for obtaining sub kJ/mol accuracy for the `bottom of the well' atomization energies, as W3.2lite is the only method that achieves this goal. W2.2h and W1h are capable of sub kcal/mol predictions; and G4, G4(MP2), and CBS-APNO afford $\sim$1 kcal/mol accuracy. For organic systems containing many hydrogens the zero-point corrections may be considered the `Achilles' Heel' of nonempirical thermochemical methods, since the error due to neglect of explicit anharmonicity is on the order of one kcal/mol. Accordingly, the RMSDs of the nonempirical W$n$ methods increase by 0.6 kcal/mol when zero-point corrections are included (specifically, the RMSDs are 1.7, 0.9, and 0.8 kcal/mol for W1h, W2.2h, and W3.2lite, respectively). The performance of the empirical G$n$ methods (G2, G3B3, G3, G4(MP2), and G4), on the other hand, improves upon inclusion of ZPVE corrections (with RMSD ranging between 0.4--0.6 kcal/mol), presumably due to the fact that their empirical correction terms were fitted against experimental atomization energies at 0 K.

The performance of different DFT exchange-correlation functionals in predicting atomization, isomerization and isodesmic reaction energies was evaluated. The atomization reactions are clearly the most daunting test; taking dispersion corrections into account  three best performers emerge: PW6B95, B2K-PLYP, and B2GP-PLYP, all of which attain a near-zero RMSD of 0.2 kcal/mol. For the isodesmic reactions the three said best performers attain RMSD of 1.7, 0.6, and 0.5 kcal/mol, respectively. And for the isomerization reactions (for which almost all the functionals perform exceptionally well) said best functionals attain RMSD of 0.3, 0.1, and 0.2 kcal/mol, respectively.

\section*{Acknowledgments}
Research at Weizmann was funded by the Israel Science Foundation (grant 709/05), the Helen and Martin Kimmel Center for Molecular Design, and the Weizmann Alternative Energy Research Initiative (AERI).
JMLM is the incumbent of the Baroness Thatcher Professorial Chair of Chemistry and a member {\em ad personam} of the Lise Meitner-Minerva Center for Computational Quantum Chemistry. 

\section*{Supporting Information}
Diagnostics for the importance of nondynamical correlation (Table S1); component breakdown of W1h atomization energies and from hypohomodesmotic and isodesmic reactions (Tables S2 and S3); basis set convergence of hypohomodesmotic and isodesmic reaction energy components (Table S4); root mean square deviations for the S22 weak interaction reference set for various DFT functionals (Table S5); component breakdown of computed enthalpy functions H$_{298}-$H$_0$ (Table S6); errors of DFT methods for individual atomization energies, isomerization reactions, and isodesmic reactions (Tables S7 through S9); B3LYP/pc-2 (and, where available, CCSD(T)/cc-pVQZ) optimized geometries in Cartesian coordinates.

\clearpage
\squeezetable
\begin{table}
\caption{Component breakdown of the final W4 total atomization energies at the bottom of the well (TAE$_e$) and comparison between W4 total atomization energies at 0 K (TAE$_0$), Active Thermochemical Tables benchmarks, and CCCBDB reference data (in kcal/mol)$^a$\label{tab:W4comp}}
\resizebox{1.1\textwidth}{!}{%
\begin{tabular}{l|ccccccccccccccc|ccc}
\hline\hline
 & SCF & CCSD & (T) & $\hat{T}_3-$(T) & $\hat{T}_4$ & $\hat{T}_5$ & inner & relativ. & spin & DBOC & $(a)$ &M-A$^b$ & TAE$_e$ & ZPVE$^c$ & TAE$_0$$^{d}$ & ATcT$^{d,e}$ & CCCBDB$^f$ & CCCBDB$^g$\\
 & & & & & & & shell &  & orbit &  & &  & &  \\
\hline
methane  & 331.55 & 84.72  & 2.89  & -0.08 & 0.08     & 0.00     & 1.27 & -0.19 & -0.08 & 0.10 & -0.05 & 0.02 & 420.21 & 27.74 & 392.47$\pm$0.16 & 392.50$\pm$0.02 & 392.51$\pm$0.07 & 392.4$\pm$0.1\\
ethane   & 558.01 & 146.34 & 6.37  & -0.35 & 0.23     & 0.01     & 2.43 & -0.39 & -0.17 & 0.14 & -0.08 & 0.04 & 712.58 & 46.39 & 666.18$\pm$0.16 & 666.16$\pm$0.05 & 666.25$\pm$0.11 & 666.1$\pm$0.2\\
ethene   & 434.97 & 119.32 & 7.40  & -0.46 & 0.43     & 0.03     & 2.38 & -0.33 & -0.17 & 0.12 & -0.07 & 0.04 & 563.64 & 31.60 & 532.04$\pm$0.16 & 532.00$\pm$0.05 & 532.06$\pm$0.13 & 531.9$\pm$0.2\\
acetylene& 299.87 & 94.73  & 8.35  & -0.72 & 0.70     & 0.07     & 2.49 & -0.28 & -0.17 & 0.12 & -0.05 & 0.04 & 405.13 & 16.46 & 388.67$\pm$0.16 & 388.67$\pm$0.05 & 388.90$\pm$0.20 & 388.7$\pm$0.3\\
propane  & 785.34 & 209.05 & 10.12 & -0.63 & 0.38$^h$ & N/A      & 3.61 & -0.58 & -0.25 & 0.20 & -0.11 & 0.06 & 1007.15 & 64.20& 942.95$\pm$0.16 & 942.78$\pm$0.10 & 942.92$\pm$0.14 & 942.7$\pm$0.3\\
propene  & 665.18 & 181.71 & 11.17 & -0.76 & 0.60$^h$ & 0.04$^i$ & 3.61 & -0.52 & -0.25 & 0.18 & -0.09 & 0.06 & 860.89 & 49.36 & 811.53$\pm$0.16 & N/A             & 811.67$\pm$0.27 & 811.4$\pm$0.4\\
propyne  & 534.14 & 155.80 & 11.94 & -1.01 & 0.87     & 0.07$^i$ & 3.76 & -0.48 & -0.25 & 0.17 & -0.07 & 0.06 & 704.97 & 34.52 & 670.45$\pm$0.16 & N/A             & 670.91$\pm$0.22 & 670.7$\pm$0.4\\
allene   & 532.55 & 155.63 & 12.27 & -0.84 & 0.74     & 0.05$^i$ & 3.67 & -0.47 & -0.25 & 0.15 & -0.07 & 0.06 & 703.45 & 34.08 & 669.37$\pm$0.16 & N/A             & 669.28$\pm$0.27 & 669.1$\pm$0.4\\
\hline\hline
\end{tabular}}
\begin{flushleft}
$(a)$$\Delta$DBOC$=$DBOC[CCSD/cc-pVDZ]$-$DBOC[HF/cc-pVDZ] correction.\\ 
$^a$W4 values for methane, ethane, ethene, and acetylene are from ref. \cite{C2H6}, and for propyne and allene from ref. \cite{B2GP-PLYP}  (except for the ZPVE corrections which were calculated in the present work, see computational details section); W4h values for propane and propene are from the present work. Note that the atomization energies include a post-SCF contributions to the diagonal Born-Oppenheimer correction, cfr. footnote $(a)$.\\
$^b$Difference between the ACES II and MOLPRO definitions of the valence ROCCSD(T), half of this contribution is added to the final TAE as discussed in the appendix of ref. \cite{w4}.\\
$^c$Methane, ethane, ethene, and acetylene from ref \cite{C2H6}; propane, propene, propyne, and allene this work (see computational details section).\\
$^d$The adjunct uncertainties correspond to 95\% confidence intervals.\\
$^e$ATcT reference values for methane, ethane, ethene, acetylene, and propane are taken from refs. \cite{Klopper09}.\\
$^f$Using ATcT atomic heats of formation at 0 K in converting the CCCBDB\cite{cccbdb} molecular heat of formation at 0 K into an atomization energy (see computational details section).\\
$^g$Using CODATA atomic heats of formation at 0 K in converting the CCCBDB\cite{cccbdb} molecular heat of formation at 0 K into an atomization energy.\\
$^h$Here $\hat{T}_4\approx 1.1 (E[\hbox{CCSDT(Q)/cc-pVTZ(no $d$ on H)}] - E[\hbox{CCSDT/cc-pVTZ(no $d$ on H)}] + E[\hbox{CCSDTQ/cc-pVDZ(no $p$ on H)}] - E[\hbox{CCSDT(Q)/cc-pVDZ(no $p$ on H)}]$.\\
$^i$The $\hat{T}_5$ contribution approximated as CCSDTQ(5)$_\Lambda$/cc-pVDZ(no $d$)$-$CCSDTQ/cc-pVDZ(no $d$).\\
\end{flushleft}
\end{table}

\clearpage

\squeezetable
\begin{table}
\caption{Component breakdown of the final W3.2lite total atomization energies and from hypohomodesmotic and isodesmic reactions (in kcal/mol).\label{tab:W3.2lite}}
\begin{tabular}{l|ccccccccccccc}
\hline\hline
 & SCF & valence & valence & $\hat{T}_3-$(T) & (Q) & inner & relativ. & spin & DBOC & M-A$^a$ & TAE$_e$ & ZPVE$^b$ & TAE$_0$\\
 &     & CCSD    & (T)     &                 &     & shell &          & orbit&      &     &         &      &        \\
\hline
methane   &331.54 & 84.76  & 2.94 &-0.12 & 0.09 & 1.28 &-0.19 & -0.08 & 0.10 & 0.02 & 420.32 & 27.57 & 392.75 \\
ethane    &558.00 & 146.40 & 6.43 &-0.33 & 0.25 & 2.45 &-0.39 & -0.17 & 0.14 & 0.04 & 712.81 & 45.98 & 666.83 \\
propane   &785.21 & 209.08 & 10.21&-0.56 & 0.43 & 3.59 &-0.58 & -0.25 & 0.20 & 0.06 & 1007.35& 63.69 & 943.67\\
$n$-butane  &1012.62& 271.72 & 13.96&-0.82 &0.61  & 4.81 &-0.77 & -0.34 & 0.25 & 0.08 & 1302.08& 81.26 & 1220.83\\
$n$-pentane &1239.85& 334.45 & 17.75&-1.07 & 0.79 & 5.99 &-0.96 & -0.44 & 0.30 & 0.10 & 1596.73& 98.88 & 1497.87\\
isobutane &1012.93& 272.63 & 14.17&-0.84 & 0.62 & 4.81 &-0.76 & -0.34 & 0.26 & 0.08 & 1303.51& 81.01 & 1222.51\\
neopentane&1240.03& 337.34 & 18.41&-1.16 & 0.82 & 5.97 &-0.95 & -0.42 & 0.32 & 0.10 & 1600.41& 98.56 & 1501.86\\
isobutene & 895.34& 244.67 & 15.09& -0.89& 0.74 & 4.92 & -0.71& -0.34 & 0.24 & 0.08 & 1159.10& 66.44 & 1092.67\\
\hline
\multicolumn{14}{c}{Hypohomodesmotic reactions, eq. \ref{reac:hypohomodesmotic}}\\
\hline
$n$-butane & 0.20 &-0.03 &-0.03 &-0.01 & 0.00 & 0.07 & 0.00 & [0] & 0.00 & [0] &0.19 & 0.13 & 0.32\\  
$n$-pentane& 0.21 & 0.02 &-0.02 &-0.03 & 0.01 & 0.11 & 0.00 & [0] &-0.01 & [0] &0.30 & 0.22 & 0.52\\  
\hline
\multicolumn{14}{c}{Isodesmic reactions, eq. \ref{reac:hypohomodesmotic}}\\
\hline
isobutane & 0.51 & 0.88 & 0.18 & -0.03 & 0.01 & 0.07 & 0.01 & [0] & 0.00 & [0] & 1.62 & 0.39 & 2.01\\  
neopentane& 0.39 & 2.90 & 0.64 & -0.11 & 0.04 & 0.09 & 0.02 & [0] & 0.02 & [0] & 3.99 & 0.53 & 4.52\\  
\hline
\multicolumn{14}{c}{Isodesmic reactions, eq. \ref{reac:isodesmic}}\\
\hline
isobutane  & 2.00 & 2.96 & 0.75 & -0.09 & 0.04 & 0.03 & 0.02 & [0] & 0.03 & [0] & 5.73 & 1.80 & 7.53\\
neopentane & 2.63 & 6.03 & 1.49 & -0.20 & 0.08 & 0.03 & 0.03 & [0] & 0.05 & [0] & 10.15& 2.65 & 12.80\\ 
\hline
\multicolumn{14}{c}{Isodesmic reactions balance the number of 1,3-interactions on both sides}\\
\hline
isobutane+3C$_2$H$_6$$\rightarrow$CH$_4$+3C$_3$H$_8$   & -0.24 & -0.17 & -0.11 & 0.00 & 0.00 & 0.09 & 0.00 & [0] & -0.01 & [0] & -0.43 & -0.32 & -0.76\\
neopentane+8C$_2$H$_6$$\rightarrow$3CH$_4$+6C$_3$H$_8$ & -1.85 & -0.23 & -0.22 &-0.02 &-0.01 & 0.15 & 0.00 & [0] & -0.02 & [0] & -2.18 & -1.59 & -3.77\\
\hline
\multicolumn{14}{c}{Isodesmic reactions balance the number of CH$_3$ groups on both sides}\\
\hline
2isobutane+CH$_4$$\rightarrow$3C$_3$H$_8$ & 1.76 & 2.79 & 0.64 & -0.10 & 0.04 & 0.12 & 0.02 & [0] & 0.02 & [0] & 5.30 & 1.48 & 6.78\\
neopentane+CH$_4$$\rightarrow$2C$_3$H$_8$ & 1.14 & 3.94 & 0.92 & -0.14 & 0.05 & 0.07 & 0.02 & [0] & 0.03 & [0] & 6.04 & 1.24 & 7.28\\
isobutene+CH$_4$$\rightarrow$C$_2$H$_4$+C$_3$H$_8$ & 6.47 & 1.25 & 0.46 & -0.09 & 0.07 & 0.15 & 0.00 & [0] & 0.03 & [0] & 8.34 & 1.19 & 9.53\\
\hline\hline
\end{tabular}
\begin{flushleft}
$^a$The difference between the ACES II and MOLPRO definitions of the valence ROCCSD(T).\\
$^b$B3LYP/pc-2 harmonic frequencies scaled by 0.985\\
\end{flushleft}
\end{table}
\clearpage

\squeezetable
\begin{table}
\caption{Total atomization energies at 0 K and heats of formation at 298 K (in kcal/mol). `Quasi-W4' atomization energies are estimated from the hypohomodesmotic reaction eq. \ref{reac:hypohomodesmotic} for linear alkanes and from the isodesmic reaction eq. \ref{reac:hypohomodesmotic} for branched alkanes.\label{tab:TAEhypo}}
\resizebox{1.1\textwidth}{!}{%
\begin{tabular}{l|cc|ccccc|cc|cccc}
\hline\hline
 & \multicolumn{2}{c}{TAE$_e$$^a$} & \multicolumn{5}{c}{TAE$_0$} & \multicolumn{2}{c}{$H_{298}-H_0$} & \multicolumn{4}{c}{$\Delta H^\circ_{f,298}$}\\
 & Quasi-W4$^b$ & Quasi-W4$^c$ & W1h & W3.2lite & Quasi-W4$^b$ & Quasi-W4$^c$ & Expt.$^d$ & Theor.$^o$ & TRC$^n$ & `Quasi-W4' & CCCBDB & \multicolumn{2}{c}{Webbook}\\
 & & & & & & & &  & &  & TRC$^m$ & Rossini$^e$ & Other\\
\hline
$n$-butane      &  1302.90 & 1302.89 & 1221.42 & 1220.83 & 1220.06 & 1220.04 & 1220.10$\pm$0.20    & 4.62 & 4.61  & -30.00  & -30.06$\pm$0.17 & -30.37$\pm$0.16 & -30.03$\pm$0.16$^g$\\
$n$-pentane     &  1597.84 & 1597.81 & 1498.61 & 1497.87 & 1497.04 & 1497.01 & 1497.33$\pm$N/A     & 5.70 & 5.78  & -34.84  & -35.09          & -35.00$\pm$0.16 & -35.16$\pm$0.24$^h$\\
                &          &         &         &         &         &         &                     &      &       &         &                 &                 & -35.08$\pm$0.14$^i$\\
$n$-hexane      &  1892.79 &         & 1775.91 &         & 1774.15 &         & 1774.28$\pm$N/A     & 6.79 & 6.86  & -39.84  & -39.89          & -39.96$\pm$0.19 & -39.94$^j$     \\
$n$-heptane     &  2187.76 &         & 2053.14 &         & 2051.17 &         & 2051.40$\pm$0.27$^f$& 7.87 & 7.94  & -44.74  &                 & -44.89$\pm$0.19 & -45.24$^k$     \\
$n$-octane      &  2482.72 &         & 2330.46 &         & 2328.30 &         & 2328.45$\pm$N/A$^f$ & 8.97 & 9.03  & -49.71  &                 & -49.82$\pm$0.16 & -49.88$^l$     \\
\hline                                                                                                    
isobutane       &  1304.35 & 1304.31 & 1223.13 & 1222.51 & 1221.76 & 1221.73 & 1221.98$\pm$0.17    & 4.29 & 4.30  & -32.01  & -32.27$\pm$0.14 & -32.42$\pm$0.13 & -32.07$\pm$0.15$^g$\\
isopentane      &  1598.74 &         & 1499.83 &         & 1498.27 &         & 1498.48$\pm$N/A     & 5.30 & 5.26  & -36.49  & -36.74          & -36.92$\pm$0.20 & -36.84$\pm$0.23$^h$\\
                &          &         &         &         &         &         &                     &      &       &         &                 &                 & -36.73$\pm$0.14$^i$\\
neopentane      &  1601.53 & 1601.46 & 1502.63 & 1501.86 & 1501.06 & 1501.01 & 1502.14$\pm$N/A     & 4.84 & 5.54  & -39.69  & -40.13          & -39.67$\pm$0.25 & -40.27$\pm$0.24$^h$\\
                &          &         &         &         &         &         &                     &      &       &         &                 &                 & -40.14$\pm$0.15$^i$\\
isohexane       &  1893.67 &         & 1776.99 &         & 1775.22 &         & 1775.53$\pm$N/A     & 6.28 & 6.29  & -41.42  & -41.73          & -41.66$\pm$0.25 &                \\
3-methylpentane &  1893.02 &         & 1776.38 &         & 1774.61 &         & 1774.86$\pm$N/A     & 6.37 & 6.23  & -40.72  & -41.11          & -41.02$\pm$0.23 &                \\
diisopropyl     &  1893.87 &         & 1777.43 &         & 1775.67 &         & 1775.41$\pm$N/A     & 6.07 & 5.85  & -42.08  & -42.04          & -42.49$\pm$0.24 &                \\
neohexane       &  1895.28 &         & 1779.04 &         & 1777.27 &         & 1777.49$\pm$N/A     & 5.98 & 6.01  & -43.77  & -43.98          & -44.35$\pm$0.23 &                \\
isoheptane      &  2188.66 &         & 2054.40 &         & 2052.43 &         & 2052.57$\pm$N/A$^f$ & 7.44 & 7.39  & -46.43  &                 & -46.60$\pm$0.30 & -46.89$^k$     \\
neoheptane      &  2190.19 &         & 2056.38 &         & 2054.41 &         &                     & 7.01 & 6.98  & -48.84  &                 & -49.29$\pm$0.32 & -50.07$^k$     \\
hexamethylethane&  2484.09 &         & 2332.98 &         & 2330.81 &         &                     & 7.52 & 7.53  & -53.68  &                 & -53.99$\pm$0.46 & -54.06$^l$     \\
isooctane       &  2484.26 &         & 2332.84 &         & 2330.67 &         &2330.86$\pm$0.37$^f$ & 7.77 & 7.69  & -53.29  &                 & -53.57$\pm$0.32 & \\
\hline\hline
\end{tabular}}
\begin{flushleft}
$^a$W4 zero-point exclusive, nonrelativistic, clamped-nuclei atomization energies for testing/parametrization of DFT functionals.
$^b$Using W1h reaction energies and W4 values for ethane and propane.
$^c$Using W3.2lite reaction energies and W4 values for ethane and propane.
$^d$$n$-octane and isooctane from ref. \cite{NIST}, all the rest from the experimental data section of ref. \cite{cccbdb}; using ATcT atomic heats of formation at 0 K in converting the molecular heat of formation at 0 K into an atomization energy (see computational details section).
$^e$$n$-butane and isobutane data were instead taken from E. J. Prosen, F. W. Maron, and F. D. Rossini, {\em J. Res. NBS} {\bf 46}, 106 (1951), higher alkanes from E. J. Prosen and F. D. Rossini, {\em J. Res. NBS} {\bf 38}, 263 (1945).
$^f$Heats of formation at 298 K were converted to 0 K using the TRC enthalpy functions, $H_{298}-H_0$, for the molecules (see table) and the CODATA\cite{codata} enthalpy functions for the elemental reference states: $H_{298}-H_0$[H$_2$(g)]=2.024$\pm$0.000, and $H_{298}-H_0$[C(cr,graphite)]=0.251$\pm$0.005 kcal/mol.
$^g$D. A. Pittam  and G. Pilcher J. Chem. Soc. Faraday Trans. 68, 2224-2229 (J. Chem. Soc. Faraday Trans. 1, 1972, 68, 2224-2229).
$^h$G. Pilcher and J. D. M. Chadwick, Trans. Faraday Soc. 63, 2357-2361 (1967).
$^i$W. D. Good, J. Chem. Thermodyn. 2, 237-244 (1970).
$^j$W. D. Good and N. K. Smith J. Chem Eng. Data 14, 102-106 (1969); liquid combined with Rossini vaporization  enthalphy (D. R. Burgess).
$^k$G. F. Davis and E. C. Gilbert J. Am. Chem. Soc. 63, 2730-2732 (1941); liquid combined with Rossini vaporization  enthalphy (D. R. Burgess).
$^l$W. D. Good, J. Chem. Thermodyn. 4, 709-714 (1972); liquid combined with Rossini vaporization  enthalphy (D. R. Burgess).
$^m$TRC (Thermodynamic Research Center) database\cite{TRC}.
$^n$Ref.\cite{TRC} All values except for $n$-butane appear to have been taken from Ref.\cite{BOM666}. For $n$-butane, the latter source lists 4.71 kcal/mol.
$^o$The translational, rotational, and vibrational contributions were obtained within the RRHO (rigid rotor-harmonic oscillator) approximation from the B3LYP/pc-2 calculated geometry and harmonic frequencies. Internal rotation corrections were obtained using the Ayala-Schlegel method\cite{AyalaSchlegel}, again on the B3LYP/pc-2 potential surface. Conformer corrections were taken from a recent benchmark study\cite{Conformers}. The specific data given here were obtained at the B2K-PLYP-D/pc-2//PW6B95/6-311G** level. See Computational Methods section for further details.
\end{flushleft}
\end{table}
\clearpage

\squeezetable
\begin{table}
\caption{Root mean square deviations (relative to our best values, in kcal/mol) of popular compound thermochemistry methods for the atomization and isomerization energies of the alkanes considered in the present work.\label{tab:Gn}}
\begin{tabular}{l|ccc|ccc}
\hline\hline
&\multicolumn{3}{c}{zero-point exclusive} & \multicolumn{3}{c}{zero-point inclusive}\\
 & Large set$^a$ & Small set$^b$ & Isomerization$^c$ & Large set$^a$ & Small set$^b$ & Isomerization$^c$\\ 
\hline
G1          & 11.21 & 7.61 & 1.19 & 7.56 & 4.98  & 0.97 \\
G2(MP2)     & 5.87  & 3.90 & 1.17 & 2.24 & 1.31  & 0.96 \\
G2          & 4.27  & 2.82 & 1.16 & 0.77 & 0.47  & 0.94 \\
G3B3        & 3.16  & 2.13 & 0.94 & 0.73 & 0.46  & 0.86 \\
G3          & 3.73  & 2.57 & 1.05 & 0.45 & 0.36  & 0.83 \\
G4(MP2)     & 1.31  & 1.03 & 1.00 & 0.77 & 0.61  & 1.01 \\
G4          & 1.29  & 0.95 & 0.93 & 0.74 & 0.53  & 0.95 \\
CBS-QB3     & 2.78  & 1.76 & 0.79 & 2.50 & 1.53  & 0.79 \\
CBS-APNO$^d$& 1.83  & 1.17 & 0.71 & 3.02 & 2.23  & 0.60 \\
W1h         & 0.94  & 0.73 &      & 1.71 & 1.28  &      \\
W2.2h       &       & 0.32 &      &      & 0.91  &      \\
W3.2lite    &       & 0.18 &      &      & 0.76  &      \\
\hline\hline
\end{tabular}
\begin{flushleft}
$^a$Includes all the C$_1$--C$_8$ alkanes considered in the present work.\\
$^b$Includes all the C$_1$--C$_5$ alkanes (w/o isopentane).\\
$^c$For the linear$\rightarrow$branched isomerization reactions.\\
$^d$Not including isooctane, for which the calculation was deemed too demanding in computer time.\\
\end{flushleft}
\end{table}
\clearpage

\squeezetable
\begin{table}
\caption{Performance statistics (in kcal/mol) of various exchange-correlation functionals with and without the $s_6$ correction for the atomization energies of the C$_1$--C$_8$ alkanes considered in the present work.\label{tab:DFTatomization}}
\begin{tabular}{l|l|l|ccc|ccc|cc}
\hline\hline
 & & &\multicolumn{3}{c}{uncorrected} & \multicolumn{5}{c}{corrected}\\
 Functional & Basis set & $s_6$ & RMSD & MSD & MAD & RMSD & MSD & MAD & $s_{\rm 6,opt}$ & RMSD$_{\rm opt}$\\
\hline
SVWN5     & pc-2     & -0.25& 231   & 218   & 218   & 227   & 215   & 215   & -16   & 44\\
\hline
PBE       & pc-2     & 0.75 & 13.70 & 12.65 & 12.71 & 24.07 & 22.08 & 22.10 & -0.93 & 3.75\\
HCTH407   & pc-2     & 1.10 & 23.95 &-21.90 & 21.90 & 8.49  & -8.07 & 8.07  & 1.69  & 1.50\\
BLYP      & pc-2     & 1.20 & 36.70 &-33.86 & 33.86 & 19.93 &-18.77 & 18.77 & 2.58  & 4.03\\
\hline
TPSS      & pc-2     & 1.00 & 4.22  & 3.56  & 3.91  & 16.81 & 16.13 & 16.13 & -0.16 & 3.53\\
M06-L      & pc-2     & 0.20 & 12.27 &-11.58 & 11.58 & 9.50  & -9.06 & 9.06  & 0.86  & 2.05\\
\hline
PBE0      & pc-2     & 0.70 & 1.70  & 0.55  & 1.29  & 10.57 & 9.35  & 9.57  & -0.04 & 1.61\\
B3PW91    & pc-2     & 1.10 & 7.02  & -6.00 & 6.00  & 8.63  & 7.83  & 7.85  & 0.49  & 0.90\\
X3LYP     & pc-2     & 0.85 & 13.53 &-11.91 & 11.95 & 1.57  & -1.22 & 1.39  & 0.96  & 0.43\\
B97-3      & pc-2     & 0.90 & 15.63 &-14.29 & 14.29 & 3.03  & -2.97 & 2.97  & 1.10  & 0.91\\
B3LYP     & pc-2     & 1.05 & 16.60 &-14.69 & 14.74 & 1.86  & -1.49 & 1.67  & 1.17  & 0.64\\
B97-2      & pc-2     & 1.05 & 17.03 &-15.77 & 15.77 & 2.59  & -2.57 & 2.57  & 1.20  & 1.48\\
B97-1      & pc-2     & 0.65 & 20.41 &-14.29 & 14.29 & 11.38 & -2.97 & 2.97  & 1.43  & 2.85\\
\hline
TPSSh     & pc-2     & 0.90 & 3.69  & 2.82  & 3.38  & 14.66 & 14.14 & 14.14 & -0.11 & 3.34\\
TPSS1KCIS & pc-2     & 0.90 & 6.33  & -5.02 & 5.18  & 6.64  & 6.29  & 6.29  & 0.44  & 1.28\\        
PW6B95    & pc-2     & 0.50 & 7.03  & -6.27 & 6.27  & 0.21  & 0.02  & 0.15  & 0.50  & 0.21\\
BMK       & pc-2     & 0.65 & 7.39  & -6.82 & 6.82  & 1.94  & 1.36  & 1.54  & 0.52  & 0.65\\
M06-2X     & pc-2     & 0.06 & 7.42  & -7.13 & 7.13  & 6.59  & -6.37 & 6.37  & 0.51  & 1.63\\
M06       & pc-2     & 0.25 & 7.56  & -7.12 & 7.12  & 4.16  & -3.97 & 3.97  & 0.52  & 1.51\\
B1B95     & pc-2     & 0.75 & 12.87 &-11.89 & 11.89 & 2.52  & -2.46 & 2.46  & 0.91  & 1.18\\
\hline
B2K-PLYP  & pc-3$^a$ & 0.30 & 4.16  & -3.78 & 3.78  & 0.22  & -0.01 & 0.16  & 0.29  & 0.20\\
B2GP-PLYP & pc-3$^a$ & 0.40 & 5.51  & -4.95 & 4.95  & 0.20  & 0.08  & 0.14  & 0.39  & 0.13\\
B2T-PLYP  & pc-3$^a$ & 0.48 & 7.99  & -7.22 & 7.22  & 1.25  & -1.18 & 1.18  & 0.56  & 0.37\\
B2-PLYP   & pc-3$^a$ & 0.55 & 8.71  & -7.81 & 7.81  & 0.98  & -0.90 & 0.90  & 0.62  & 0.31\\        
mPW2-PLYP & pc-3$^a$ & 0.40 & 12.48 &-11.39 & 11.39 & 6.86  & -6.36 & 6.36  & 0.88  & 0.96\\
B2K-PLYP  & pc-2$^b$ & 0.30 & 5.79  & -5.36 & 5.36  & 1.62  & -1.59 & 1.59  & 0.41  & 0.57\\
B2GP-PLYP & pc-2$^b$ & 0.40 & 7.13  & -6.52 & 6.52  & 1.54  & -1.49 & 1.49  & 0.50  & 0.52\\
B2T-PLYP  & pc-2$^b$ & 0.48 & 9.59  & -8.76 & 8.76  & 2.87  & -2.73 & 2.73  & 0.68  & 0.75\\
B2-PLYP   & pc-2$^b$ & 0.55 & 10.35 & -9.40 & 9.40  & 2.64  & -2.49 & 2.49  & 0.73  & 0.69\\        
mPW2-PLYP & pc-2$^b$ & 0.40 & 14.04 &-12.88 & 12.88 & 8.43  & -7.86 & 7.86  & 0.99  & 1.33\\                
\hline
MP2       & pc-3     & -0.15& 4.08  & -3.99 & 3.99  & 5.89  & -5.88 & 5.88  & 0.23  & 2.56\\
MP2-SCS   & pc-3     & 0.17 & 2.78  & -2.63 & 2.63  & 0.56  & -0.49 & 0.53  & 0.19  & 0.44\\
MP2       & pc-2     & -0.15& 18.42 &-17.97 & 17.97 & 20.46 &-19.85 & 19.85 & 1.25  & 5.43\\
MP2-SCS   & pc-2     & 0.17 & 19.90 &-18.88 & 18.88 & 17.55 &-16.74 & 16.74 & 1.38  & 3.98\\
\hline\hline
\end{tabular}
\begin{flushleft}
$^a$(frozen core) pc-3 basis set combined with a CBS extrapolation where Nmin=15 as recommended in Ref. \cite{Petersson}.\\
$^b$(frozen core) pc-2 basis set combined with a CBS extrapolation where Nmin=10 as recommended in Ref. \cite{Petersson}.\\
\end{flushleft}
\end{table}
\clearpage

\squeezetable
\begin{table}
\caption{Performance statistics (in kcal/mol) of various exchange-correlation functionals with and without the $s_6$ correction for the reaction energies of the isomerization linear-alkane$\rightarrow$branched-alkane for the C$_4$--C$_8$ alkanes considered in the present work.\label{tab:DFTisomerization}}
\begin{tabular}{l|l|l|ccc|ccc}
\hline\hline
 & & &\multicolumn{3}{c}{uncorrected} & \multicolumn{3}{c}{corrected}\\
 Functional & Basis set & $s_6$ & RMSD & MSD & MAD & RMSD & MSD & MAD \\
\hline
SVWN5     & pc-2     & -0.25 & 0.67 & -0.56 & 0.56 & 0.44 & 0.30  & 0.30\\
\hline
PBE       & pc-2     & 0.75  & 2.90 & 2.34  & 2.34 & 0.36 & -0.25 & 0.28\\
HCTH407   & pc-2     & 1.10  & 5.39 & 4.26  & 4.26 & 0.68 & 0.46  & 0.46\\
BLYP      & pc-2     & 1.20  & 4.52 & 3.65  & 3.65 & 0.68 & -0.50 & 0.51\\
\hline    
TPSS      & pc-2     & 1.00  & 3.50 & 2.90  & 2.90 & 0.87 & -0.55 & 0.60\\
M06-L      & pc-2     & 0.20  & 1.07 & 0.94  & 0.94 & 0.48 & 0.25  & 0.40\\
\hline    
PBE0      & pc-2     & 0.70  & 2.79 & 2.26  & 2.26 & 0.26 & -0.15 & 0.19\\
B3PW91    & pc-2     & 1.10  & 3.82 & 3.11  & 3.11 & 0.95 & -0.68 & 0.69\\
X3LYP     & pc-2     & 0.85  & 3.65 & 2.94  & 2.94 & 0.13 & 0.00  & 0.10\\
B97-3      & pc-2     & 0.90  & 3.89 & 3.15  & 3.15 & 0.17 & 0.04  & 0.12\\
B3LYP     & pc-2     & 1.05  & 3.99 & 3.22  & 3.22 & 0.56 & -0.41 & 0.42\\
B97-2      & pc-2     & 1.05  & 4.18 & 3.38  & 3.38 & 0.38 & -0.24 & 0.29\\
B97-1      & pc-2     & 0.65  & 3.09 & 2.51  & 2.51 & 0.33 & 0.26  & 0.28\\
\hline    
TPSSh     & pc-2     & 0.90  & 3.42 & 2.83  & 2.83 & 0.55 & -0.28 & 0.36\\
TPSS1KCIS & pc-2     & 0.90  & 3.74 & 3.06  & 3.06 & 0.27 & -0.04 & 0.19\\
PW6B95    & pc-2     & 0.50  & 2.09 & 1.76  & 1.76 & 0.28 & 0.03  & 0.20\\
BMK       & pc-2     & 0.65  & 1.52 & 1.27  & 1.27 & 1.31 & -0.98 & 0.98\\
M06-2X     & pc-2     & 0.06  & 0.21 & 0.19  & 0.19 & 0.15 & -0.02 & 0.10\\
M06       & pc-2     & 0.25  & 0.26 & -0.15 & 0.18 & 1.31 & -1.01 & 1.01\\
B1B95     & pc-2     & 0.75  & 2.47 & 2.09  & 2.09 & 0.83 & -0.50 & 0.54\\
\hline
B2K-PLYP  & pc-3$^a$ & 0.30  & 1.20 & 0.96  & 0.96 & 0.11 & -0.08 & 0.08\\
B2GP-PLYP & pc-3$^a$ & 0.40  & 1.52 & 1.22  & 1.22 & 0.21 & -0.16 & 0.16\\
B2T-PLYP  & pc-3$^a$ & 0.48  & 1.80 & 1.45  & 1.45 & 0.27 & -0.21 & 0.21\\
B2-PLYP   & pc-3$^a$ & 0.55  & 2.06 & 1.66  & 1.66 & 0.32 & -0.24 & 0.24\\
mPW2-PLYP & pc-3$^a$ & 0.40  & 2.15 & 1.73  & 1.73 & 0.43 &  0.35 & 0.35\\
B2K-PLYP  & pc-2$^b$ & 0.30  & 1.03 & 0.84 & 0.84  & 0.27 & -0.20 & 0.20\\
B2GP-PLYP & pc-2$^b$ & 0.40  & 1.37 & 1.11 & 1.11  & 0.35 & -0.27 & 0.27\\
B2T-PLYP  & pc-2$^b$ & 0.48  & 1.67 & 1.35 & 1.35  & 0.40 & -0.31 & 0.31\\
B2-PLYP   & pc-2$^b$ & 0.55  & 1.94 & 1.57 & 1.57  & 0.43 & -0.32 & 0.32\\
mPW2-PLYP & pc-2$^b$ & 0.40  & 2.04 & 1.65 & 1.65  & 0.34 & 0.27  & 0.27\\
\hline    
MP2       & pc-3     & -0.15 & 1.12 & -0.90 & 0.90 & 0.48 & -0.38 & 0.38\\
MP2-SCS   & pc-3     & 0.17  & 0.28 & 0.23 & 0.23  & 0.47 & -0.35 & 0.35\\
MP2       & pc-2     & -0.15 & 1.24 & -0.97 & 0.97 & 0.60 & -0.45 & 0.45\\
MP2-SCS   & pc-2     & 0.17  & 0.13 & 0.11 & 0.11  & 0.67 & -0.48 & 0.48\\
\hline\hline
\end{tabular}
\begin{flushleft}
$^a$(frozen core) pc-3 basis set combined with a CBS extrapolation where Nmin=15 as recommended in Ref. \cite{Petersson}.\\
$^b$(frozen core) pc-2 basis set combined with a CBS extrapolation where Nmin=10 as recommended in Ref. \cite{Petersson}.\\
\end{flushleft}
\end{table}
\clearpage

\squeezetable
\begin{table}
\caption{Performance statistics (RMSD, kcal/mol) of various exchange-correlation functionals with and without the $s_6$ correction for the isodesmic reaction \ref{reac:isodesmic} for the C$_3$--C$_8$ alkanes considered in the present work.\label{tab:DFTisodesmic}}
\begin{tabular}{l|l|l|cccc|cccc}
\hline\hline
 &  &  & \multicolumn{4}{c}{uncorrected} & \multicolumn{4}{c}{corrected}\\
 &  &  & $n$-alkane  & iso-alkane  & neo-alkane & all & $n$-alkane  & iso-alkane  & neo-alkane & all\\
Functional & Basis set & $s_6$ & (6 species) & (4 species) &  (3 species) &  (13 species) & (6 species) & (4 species) &  (3 species) &  (13 species)\\
\hline
VWN5      & pc-2     & -0.25 & 0.38 & 0.80 & 0.90 & 0.68 & 0.61 & 0.98 & 1.03 & 0.86\\ 
PBE       & pc-2     & 0.75  & 3.35 & 5.53 & 6.12 & 4.88 & 0.46 & 0.31 & 0.39 & 0.40\\ 
HCTH407   & pc-2     & 1.10  & 4.42 & 8.18 & 9.16 & 7.09 & 0.19 & 0.49 & 0.72 & 0.46\\ 
BLYP      & pc-2     & 1.20  & 4.99 & 8.34 & 9.24 & 7.33 & 0.37 & 0.22 & 0.17 & 0.29\\ 
TPSS      & pc-2     & 1.00  & 4.94 & 7.74 & 8.47 & 6.87 & 1.09 & 0.84 & 0.84 & 0.96\\ 
M06-L      & pc-2     & 0.20  & 3.29 & 4.45 & 4.75 & 4.07 & 2.52 & 3.07 & 3.22 & 2.88\\ 
PBE0      & pc-2     & 0.70  & 3.39 & 5.55 & 6.03 & 4.87 & 0.69 & 0.67 & 0.67 & 0.68\\ 
B3PW91    & pc-2     & 1.10  & 4.53 & 7.48 & 8.16 & 6.56 & 0.29 & 0.32 & 0.33 & 0.31\\ 
X3LYP     & pc-2     & 0.85  & 4.13 & 6.87 & 7.54 & 6.03 & 0.86 & 0.93 & 1.03 & 0.93\\ 
B97-3      & pc-2     & 0.90  & 4.65 & 7.60 & 8.32 & 6.69 & 1.19 & 1.32 & 1.43 & 1.29\\ 
B3LYP     & pc-2     & 1.05  & 4.52 & 7.52 & 8.26 & 6.60 & 0.48 & 0.25 & 0.25 & 0.37\\ 
B97-2      & pc-2     & 1.05  & 4.65 & 7.81 & 8.55 & 6.83 & 0.61 & 0.50 & 0.52 & 0.55\\ 
B97-1      & pc-2     & 0.65  & 3.93 & 6.30 & 6.90 & 5.57 & 1.43 & 1.76 & 1.92 & 1.67\\ 
TPSSh     & pc-2     & 0.90  & 4.79 & 7.54 & 8.21 & 6.68 & 1.33 & 1.29 & 1.33 & 1.31\\ 
TPSS1KCIS & pc-2     & 0.90  & 4.76 & 7.67 & 8.42 & 6.78 & 1.30 & 1.40 & 1.53 & 1.39\\ 
PW6B95    & pc-2     & 0.50  & 3.43 & 5.18 & 5.66 & 4.64 & 1.50 & 1.70 & 1.84 & 1.65\\ 
BMK       & pc-2     & 0.65  & 3.34 & 4.69 & 4.92 & 4.22 & 0.84 & 0.44 & 0.29 & 0.63\\ 
M06-2X     & pc-2     & 0.06  & 1.35 & 1.71 & 1.62 & 1.55 & 1.12 & 1.29 & 1.16 & 1.19\\ 
M06       & pc-2     & 0.25  & 1.74 & 1.88 & 1.52 & 1.75 & 0.78 & 0.41 & 0.44 & 0.60\\ 
B1B95     & pc-2     & 0.75  & 3.86 & 5.93 & 6.46 & 5.28 & 0.97 & 0.76 & 0.74 & 0.85\\ 
B2K-PLYP  & pc-3$^a$ & 0.30  & 1.76 & 2.73 & 2.82 & 2.39 & 0.61 & 0.64 & 0.52 & 0.60\\ 
B2GP-PLYP & pc-3$^a$ & 0.40  & 2.08 & 3.28 & 3.44 & 2.87 & 0.54 & 0.49 & 0.38 & 0.49\\ 
B2T-PLYP  & pc-3$^a$ & 0.48  & 2.36 & 3.76 & 3.99 & 3.30 & 0.51 & 0.42 & 0.32 & 0.44\\ 
B2-PLYP   & pc-3$^a$ & 0.55  & 2.60 & 4.19 & 4.49 & 3.68 & 0.49 & 0.37 & 0.28 & 0.41\\ 
mPW2-PLYP & pc-3$^a$ & 0.40  & 2.71 & 4.37 & 4.68 & 3.83 & 1.17 & 1.58 & 1.62 & 1.43\\
B2K-PLYP  & pc-2$^b$ & 0.30  & 1.79 & 2.65 & 2.74 & 2.34 & 0.64 & 0.56 & 0.44 & 0.57\\ 
B2GP-PLYP & pc-2$^b$ & 0.40  & 2.11 & 3.21 & 3.38 & 2.83 & 0.57 & 0.44 & 0.32 & 0.48\\ 
B2T-PLYP  & pc-2$^b$ & 0.48  & 2.39 & 3.70 & 3.94 & 3.27 & 0.54 & 0.38 & 0.27 & 0.44\\ 
B2-PLYP   & pc-2$^b$ & 0.55  & 2.64 & 4.15 & 4.46 & 3.66 & 0.53 & 0.35 & 0.25 & 0.42\\ 
mPW2-PLYP & pc-2$^b$ & 0.40  & 2.73 & 4.32 & 4.64 & 3.80 & 1.19 & 1.53 & 1.57 & 1.40\\ 
MP2       & pc-3     & -0.15 & 0.51 & 1.24 & 1.66 & 1.12 & 0.08 & 0.21 & 0.52 & 0.28\\ 
MP2-SCS   & pc-3     & 0.17  & 1.18 & 1.52 & 1.38 & 1.35 & 0.53 & 0.37 & 0.10 & 0.41\\ 
MP2       & pc-2     & -0.15 & 0.24 & 1.09 & 1.37 & 0.92 & 0.34 & 0.21 & 0.23 & 0.28\\ 
MP2-SCS   & pc-2     & 0.17  & 1.36 & 1.57 & 1.50 & 1.47 & 0.71 & 0.48 & 0.19 & 0.55\\ 
\hline\hline
\end{tabular}
\begin{flushleft}
$^a$(frozen core) pc-3 basis set combined with a CBS extrapolation where Nmin=15 as recommended in Ref. \cite{Petersson}.\\
$^b$(frozen core) pc-2 basis set combined with a CBS extrapolation where Nmin=10 as recommended in Ref. \cite{Petersson}.\\
\end{flushleft}
\end{table}


\begin{thebibliography}{99}
\bibitem{RefRep} One of us (JMLM), some years back, received a referee report containing the puzzling statement: ``Recently evidence has been accumulating that popular DFT methods are not size-consistent.'' With the benefit of hindsight, this appears to have been an allusion to the problem discussed in the text.
\end{thebibliography}
\end{document}